\documentclass[12pt, draftclsnofoot, onecolumn]{IEEEtranTCOM}
\renewcommand{\baselinestretch}{1.55}

\usepackage{indentfirst}
\usepackage{multirow}
\usepackage{graphicx}
\usepackage{color}
\usepackage{xcolor}
\usepackage{colortbl} 
\usepackage{listings}
\usepackage{enumerate}
\usepackage{subfigure}
\usepackage{amsmath,amsthm,amscd,amssymb,latexsym,upref,stmaryrd}
\usepackage[noend]{algpseudocode}
\usepackage{algorithmicx,algorithm}
\usepackage{amsfonts,epsfig}
\usepackage{CJK}
\usepackage{float}
\usepackage{cite}
\usepackage{subfigure}
\usepackage{lipsum}
\usepackage{multicol}
\usepackage{setspace}
\usepackage{bm}
\usepackage{stfloats}
\usepackage[aboveskip=-0pt,belowskip=-30pt]{caption}
\usepackage{geometry}
\usepackage{booktabs}
\usepackage{threeparttable}
\newtheorem{remark}{Remark}
\usepackage[utf8]{inputenc}

\begin{document}	
\title{Model-based Learning Network for 3-D Localization in mmWave Communications}

\author{Jie~Yang,
	Shi~Jin,	Chao-Kai~Wen,
	Jiajia~Guo,
	Michail~Matthaiou, 
	and~Bo~Gao
	
	\thanks{This paper was presented in part at the IEEE VTC-Fall, Honolulu, HI, USA, September 2019 \cite{conference}. Jie~Yang, Shi~Jin, and Jiajia~Guo are with the National Mobile Communications Research Laboratory, Southeast University, Nanjing, China (e-mail: \{yangjie;jinshi;jiajiaguo\}@seu.edu.cn). Chao-Kai~Wen is with the Institute of Communications Engineering, National Sun Yat-sen University, Kaohsiung, 804, Taiwan (e-mail: chaokai.wen@mail.nsysu.edu.tw). Michail~Matthaiou is with the Institute of Electronics, Communications and Information Technology (ECIT), Queen's University Belfast, Belfast, U.K. (e-mail: m.matthaiou@qub.ac.uk).
	Bo~Gao is with the ZTE Corporation and the State Key Laboratory of Mobile Network and Mobile Multimedia Technology, Shenzhen, China (e-mail: gao.bo1@zte.com.cn).}
}

\maketitle	
\vspace{-1.5cm}
\begin{abstract}
\vspace{-0.3cm}
Millimeter-wave (mmWave) cloud radio access networks (CRANs) provide new opportunities for accurate cooperative localization, in which large bandwidths and antenna arrays and increased densities of base stations enhance the delay and angular resolution. 
This study considers the joint location and velocity estimation of user equipment (UE) and scatterers in a three-dimensional mmWave CRAN architecture. 
Several existing works have achieved satisfactory results by using neural networks (NNs) for localization.
However, the black box NN localization method has limited robustness and accuracy and relies on a prohibitive amount of training data to increase localization accuracy.
Thus, we propose a model-based learning network for localization to address these problems. 
In comparison with the black box NN, we combine NNs with geometric models.
Specifically,
we first develop an unbiased weighted least squares (WLS) estimator by utilizing hybrid delay and angular measurements, which determine the location and velocity of the UE in only one estimator, and can obtain the location and velocity of scatterers further.
The proposed estimator can achieve the Cram\'{e}r-Rao lower bound under small measurement noise and outperforms other state-of-the-art methods.
Second, we establish a NN-assisted localization method called NN-WLS by replacing the linear approximations in the proposed WLS localization model with NNs to learn the higher-order error components, 
thereby enhancing the performance of the estimator, especially in a large noise environment.
The solution possesses the powerful learning ability of the NN and the robustness of the proposed geometric model.
Moreover, the ensemble learning is applied to improve the localization accuracy further. 
Comprehensive simulations show that the proposed NN-WLS is superior to the benchmark methods in terms of localization accuracy, robustness, and required time resources.

\end{abstract}

\vspace{-0.3cm}
\begin{IEEEkeywords}
	\vspace{-0.3cm}
Cooperative localization,
cloud radio access network,
hybrid measurements,
millimeter-wave communications,
neural network,
weighted least squares.
\vspace{-0.2cm}
\end{IEEEkeywords}

\section{Introduction}
Future networks should offer unlimited coverage to any devices anywhere and anytime to stimulate the amalgamation of localization and wireless communications \cite{locsum}.	
Millimeter-wave (mmWave) communication is a promising technology for meeting such requirements in future wireless communications. Localization is a highly desirable feature of mmWave communications \cite{loc0,loc1}. The user equipment (UE) location can be used to provide location-based services, such as navigation, mapping, social networking, augmented reality, and intelligent transportation systems. Additionally, location-aware communications can be realized by the obtained location information to improve communication capacity and network efficiency \cite{loc}.

MmWave bands offer larger bandwidths than the presently used sub-6 GHz bands, hence, higher resolution of the time of arrival (TOA), time difference of arrival (TDOA), and frequency difference of arrival (FDOA) can be consequently achieved. In addition, the penetration loss from mmWave bands is inherently large \cite{overview,Han2017,M2017}. Thus, the difference between the received power of the line-of-sight (LOS) path and the non-LOS (NLOS) path is pronounced, thereby simplifying the elimination of NLOS interference \cite{sparse,path,direct4}. To compensate for severe penetration loss and increased path-loss, large antenna arrays and highly directional transmission should be combined to facilitate the acquisition of the angle of arrival (AOA) and the angle of departure (AOD) \cite{ce}. Moreover, cloud radio access networks (CRANs) can enhance mmWave communication by improving the network coverage \cite{cran}. 
CRANs provide a cost-effective way to achieve network densification, in which distributed low-complexity remote radio heads (RRHs) are deployed close to the UE and coordinated by a central unit (CU) for joint processing.
The obtained location information can be shared with network nodes.
Therefore, mmWave CRANs can offer accurate cooperative localization in urban and indoor environments, wherein conventional GPS may fail \cite{co,co1,co2}. 
Channel parameters required in localization can be measured accurately \cite{ce1,ce2,ce3,ce4,ce5} in static and mobile scenarios in the initial access and communication stages owing to the remarkable delay and angular resolution of mmWave communication systems without the need to install additional expensive infrastructure.

Localization has become a popular research topic in recent years. Different localization techniques have been summarized in \cite{loc2}.
Currently, widespread localization methods apply the principle in which the channel parameters (e.g., AOA, TOA, TDOA, and FDOA) are initially extracted from the received waveform and grouped together as a function of the location parameters, and then different estimators are used to determine the UE locations.
The classical linear weighted least squares (WLS) estimators were applied in \cite{TD1,TD3,AOA,l2,l3,l4}.
In \cite{TD1,TD3}, several closed-form TOA-based WLS estimators have been proposed.
A few AOA-based methods were developed in \cite{AOA} and in the related references.
AOA and its combination with ranging estimates are expected to achieve high location accuracy.
Reference \cite{l2} considered the localization problem of the three-dimensional (3-D) stationary targets in Multiple-Input Multiple-Output (MIMO) radar systems that utilized hybrid TOA/AOA measurements,
from which a computationally efficient closed-form algorithm was developed with the WLS estimator, to achieve the Cram\'{e}r-Rao lower bound (CRLB) under small measurement noise.
Comparison shows that less effort has been devoted to the localization of moving targets.
Reference \cite{l3} estimated location and velocity by introducing two-stage WLS estimators and using the hybrid TDOA/FDOA measurements.
Reference \cite{l4} developed a WLS estimator to estimate the location and velocity of a moving target with a constant-velocity in a two-dimensional (2-D) scenario.
Nevertheless, the aforementioned studies have overlooked the localization of scatterers.
Recently, \cite{foe} advocated that future communication systems will turn multipath channels ``from foe to friend"  by leveraging distinguishable multipath components that resulted from unparalleled delay and angular resolution in mmWave systems.
Thus, the information from reflected signals can be exploited in the reconstruction of the 3-D map of the surrounding environment.
In this study, we consider the joint location and velocity estimation of a moving UE, as well as scatterers, in the 3-D scenario with mmWave communication systems by using hybrid TDOA/FDOA/AOA measurements.
Unlike closed-form methods with multistage estimators, the proposed method determines the location and velocity of the UE in only one estimator.

All of the aforementioned localization techniques \cite{TD1,TD3,AOA,l2,l3,l4} are geometric approaches, in which delay and angular measurements are extracted and from which the location and velocity of the UE, as well as the scatterers, are triangulated or trilaterated.
A function can be approximated by geometric techniques given the existence of an underlying transfer function between the measurements and the locations.
In recent years, artificial intelligence (AI) has received considerable attention because of its promising performance in solving complicated problems.
Researchers have utilized neural networks (NNs) to learn underlying transfer functions. Meanwhile, AI-based localization solutions, such as fingerprinting methods \cite{fingerprint0,fingerprint1}, have emerged.
A deep learning-based indoor fingerprinting system was presented in \cite{fingerprint0} to achieve meter-level localization accuracy. The experiments in \cite{fingerprint1} showed the feasibility of using deep learning methods for localization in actual outdoor environments.
AI-based fingerprinting methods have alleviated modeling issues and can provide better performance than model-based localization techniques that use geometric relationships by fitting real-life measurements \cite{AI1,AI2}.
However, extremely large amounts of training data are required to meet the high requirements of localization accuracy.
Purely data-based and model-based, and hybrid data and model-based wireless network designs are discussed in \cite{DoM}.
To overcome the disadvantages of purely data- or model-based localization methods, we conceive hybrid data- and model-based localization methods
by building and enhancing our localization estimator on the geometric model with NNs.
At present, the literature on localization by combining NNs with geometric models, which is the focus of the current work, is scarce.

This study addresses the 3-D localization of moving UE and scatterers in mmWave communication systems. To our best knowledge, the present study is the first to combine the WLS estimator and NNs in 3-D localization problems. The contributions of this study are presented as follows:

\begin{itemize}
\item \textbf{Localization Model}:
First, we establish a joint location and velocity estimation model by utilizing hybrid TDOA/FDOA/AOA measurements. Then, we develop an efficient closed-form WLS estimator. Unlike other closed-form WLS-based methods \cite{l3} with multistage estimators, the proposed method can determine the UE’s location and velocity in only one stage. 
Second, we exploit the single-bounce NLOS paths and the estimated UE location and velocity to build the scatterer localization model.
Then, we deduce the closed-form WLS estimator to determine the scatterers' location and velocity. 
The proposed estimator is proven asymptotically unbiased and able to attain CRLB under small measurement noise through simulations.

\item \textbf{Learning Network}:
Although the proposed WLS estimator performs well,
its performance starts deteriorating as the noise level increases.
Therefore, we propose a NN-assisted WLS method called NN-WLS to improve the localization accuracy further. The NN-WLS benefits from the powerful learning ability of the NN and the robustness of the geometric model.
In addition, the proposed NN-WLS is fast because it can eliminate iterations in the proposed WLS algorithm.
Furthermore, we embed ensemble learning into the proposed NN-WLS method to enhance localization accuracy.
Simulation results show that NN-WLS outperforms the WLS estimator significantly when the measurement noise has an intrinsic relationship.
In addition, 
the proposed NN-WLS is superior in terms of localization accuracy and robustness based on a comprehensive comparison with benchmark methods.
\end{itemize}

{\bf Notations}---Uppercase boldface $\mathbf{A}$ and lowercase boldface $\mathbf{a}$ denote matrices and vectors, respectively. For any matrix $\mathbf{A}$, the superscripts $\mathbf{A}^{-1}$ and $\mathbf{A}^{T}$ stand for inverse and transpose, respectively.
For any vector $\mathbf{a}$, the 2-norm is denoted by $\|\mathbf{a}\|$. 
$\text{diag}\{\cdot\}$ denotes a diagonal matrix with entries in $\{\cdot\}$,
and
$\mbox{blkdiag}(\mathbf{A}_1,\ldots,\mathbf{A}_k)$ denotes a block-diagonal matrix constructed
by $\mathbf{A}_1,\ldots,\mathbf{A}_k$.
$\mathbb{E}\{\cdot\}$ denotes statistical expectation, whilst $|\cdot|$ denotes the module of a complex value or the cardinality of a set.
The notation $a^\circ$ is the true value of the estimated parameter $a$.

\section{System Model}
\begin{figure}
	\vspace{-0.8cm}
	\centering
	\includegraphics[scale=0.7,angle=0]{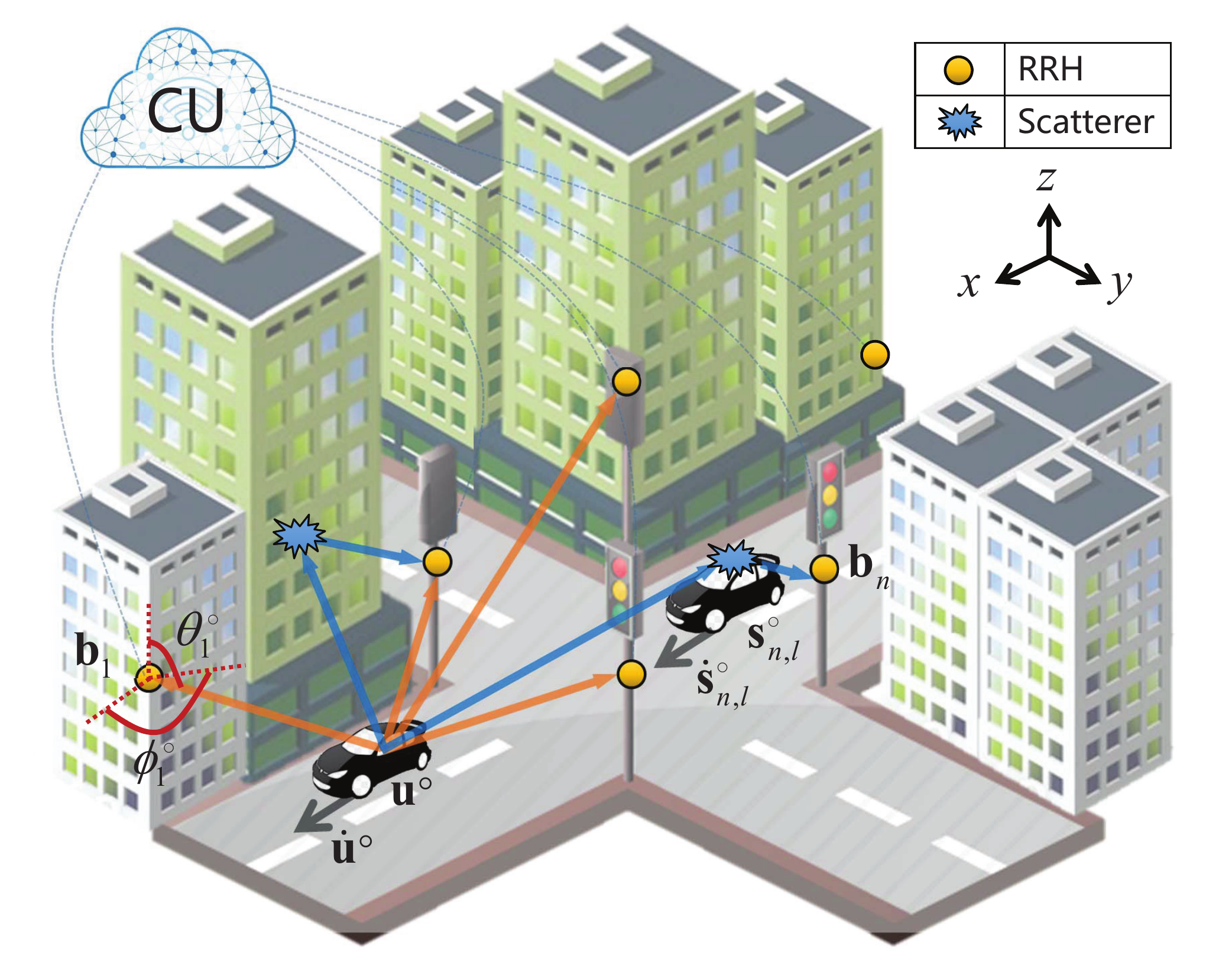}	
	\captionsetup{font=footnotesize}
	\caption{Illustration of the mmWave CRAN system model, in which RRHs are connected with the CU.}
	\label{fig:model}	
\end{figure}

We study the moving UE and scatterer localization problems in a mmWave CRAN with $N$ RRHs \cite{cran} (Fig. \ref{fig:model}).
Each RRH is equipped with a large antenna array with $K$ antenna elements and connected to the CU via an individual fronthaul link. 
We assume that the clocks of different RRHs in the CRAN are synchronized. 
For ease of expression, we consider the system model with
a single UE. The system model can be easily extended to solve the case with multiple UE as long as the pilot signals
for different UE are orthogonal in time.
The important variables are summarized in Table \ref{NOTATIONS}.

\begin{table}
	\vspace{0.8cm}
	\centering
	\caption{Notations of important variables.}\label{NOTATIONS}
	\renewcommand{\arraystretch}{1.2}
	\fontsize{8}{8}\selectfont
	\begin{tabular}{l l l l}			
		\hline		
		Notation & Definition & Notation & Definition \\
		\hline
		$\mathbf{b}_n$ & location of the $n$-th RRH & $\mathbb{M}_{n}$ & set of measurements of the $n$-th RRH\\
		\hline
		$\mathbf{u}^\circ$ & location of the UE &  $\mathbb{M}_{a}$ & set of selected LOS measurements\\
		\hline
		${\dot{\mathbf{u}}^\circ}$ & velocity of the UE& $\mathbb{M}_{r,n}$ & set of remaining measurements  of the $n$-th RRH\\ 
		\hline	
		$\mathbf{x}^\circ$ & $\mathbf{x}^\circ=[\mathbf{u}^{\circ T}, \dot{\mathbf{u}}^{\circ T}]^T$& $\phi_{n}^\circ$ & azimuth AOA-related parameter \\
		& 6-dimensional state vector of the UE & & for LOS path of the $n$-th RRH\\ 
		\hline 
		$\mathbf{s}_{n,l}^\circ$ & location of the $l$-th scatterer & $\theta_{n}^\circ$ & elevation AOA-related parameter\\
		& between the $n$-th RRH and the UE & &for LOS path of the $n$-th RRH\\ 
		\hline
		$\dot{\mathbf{s}}_{n,l}^{\circ}$ & velocity of the $l$-th scatterer & $\phi_{n,l}^{s\circ}$ & azimuth AOA-related parameter\\
		& between the $n$-th RRH and the UE & &for $l$-th NLOS path of the $n$-th RRH\\ 
		\hline
		$\mathbf{x}_{n,l}^{s\circ}$ & $\mathbf{x}_{n,l}^{s\circ} = [\mathbf{s}_{n,l}^{\circ T},\dot{\mathbf{s}}_{n,l}^{\circ T}]^T$, 6-dimensional state vector & $\theta_{n,l}^{s\circ}$ & elevation AOA-related parameter\\
		& of the $l$-th scatterer between the $n$-th RRH and the UE & & for $l$-th NLOS path of the $n$-th RRH \\ \hline
		$\alpha_{n,l}^\circ$ & complex gain for the $l$-th path of the $n$-th RRH & $\mathbf{m}$ & vector of noisy measurements in $\mathbb{M}_{a}$ \\ \hline
		$\tau_{n,l}^\circ$ & delay for the $l$-th path of the $n$-th RRH & $\mathbf{m}^\circ$ & true value of measurements corresponding to $\mathbf{m}$\\ \hline
		$\phi_{n,l}^\circ$ & azimuth AOA for the $l$-th path of the $n$-th RRH & $\Delta \mathbf{m}$ & Gaussian noise vector corresponding to $\mathbf{m}$ \\
		& & & with zero mean and covariance matrix $\mathbf{Q}$ \\ \hline
		$\theta_{n,l}^\circ$ & elevation AOA for the $l$-th path of the $n$-th RRH & $\mathbf{m}_{n,l}^{s}$ & vector of noisy measurements in $\mathbb{M}_{r,n}$\\\hline
		$\nu_{n,l}^\circ$ & Doppler shift for the $l$-th path of the $n$-th RRH & $\mathbf{m}_{n,l}^{s\circ}$ & true value of measurements corresponding to $\mathbf{m}_{n,l}^{s}$\\	\hline
		$r_{n1}^\circ$ & TDOA-related parameter & $\Delta\mathbf{m}_{n,l}^{s}$ & Gaussian noise vector corresponding to $\mathbf{m}_{n,l}^{s}$\\
		& for LOS path of the $n$-th RRH & & with zero mean and covariance matrix $\mathbf{Q}_{n,l}^{s}$\\\hline
		$r_{n1,l}^{s\circ}$ & TDOA-related parameter & $N$ & number of RRHs\\
		& for $l$-th NLOS path of the $n$-th RRH & 	& \\ \hline
		$\dot{r}_{n1}^\circ$ & FDOA-related parameter & $N_a$ & number of selected LOS paths \\
		& for LOS path of the $n$-th RRH\\ 	\hline
		$\dot{r}_{n1}^{s\circ}$ & FDOA-related parameter\\
		& for $l$-th NLOS path of the $n$-th RRH\\ 			
		\hline			
	\end{tabular}
\end{table}

\vspace{-0.25cm}
\subsection{System Geometry}
We consider a 3-D space $\mathbb{R}^3=\{[x, y, z]^T: x, y, z\in\mathbb{R}\}$ with $N$ known RRHs located at $\mathbf{b}_n=[x_n^b, y_n^b, z_n^b]^T$, for $n=1, 2, \ldots, N$.
The geometry between the RRHs and the UE is shown in Fig. \ref{fig:model}.
We assume that the unknown location and velocity of the UE
are represented by $\mathbf{u}^\circ = [x^\circ, y^\circ, z^\circ]^T$ and ${\dot{\mathbf{u}}^\circ}=[\dot{x}^\circ, \dot{y}^\circ, \dot{z}^\circ]^T$, respectively.
Note that $\mathbf{u}^\circ$ is a function of time with ${\partial \mathbf{u}^\circ }/{\partial t} = {\dot{\mathbf{u}}^\circ}$.
We only consider the LOS and the single-bounce NLOS paths
because of the sparsity and high path loss of the mmWave channel\cite{sparse,path}.
The unknown location and velocity of the $l$-th scatterer between the $n$-th RRH and the UE are represented by $\mathbf{s}_{n,l}^\circ = [x_{n,l}^{s\circ}, y_{n,l}^{s\circ}, z_{n,l}^{s\circ}]^T$ and $\dot{\mathbf{s}}_{n,l}^{\circ} = [\dot{x}_{n,l}^{s\circ}, \dot{y}_{n,l}^{s\circ}, \dot{z}_{n,l}^{s\circ}]^T$, respectively, with $l=1, 2, \ldots, L_{n}$, where $L_{n}$ is the number of scatterers between the $n$-th RRH and the UE.
Here, $\mathbf{s}_{n,l}^\circ$ is a function of time with ${\partial \mathbf{s}_{n,l}^\circ }/{\partial t} = \dot{\mathbf{s}}_{n,l}^{\circ}$.
We aim to determine $\mathbf{u}^{\circ}$, $\dot{\mathbf{u}}^{\circ}$, $\mathbf{s}_{n,l}^{\circ}$, and $\dot{\mathbf{s}}_{n,l}^{\circ}$, where $l=1, 2, \ldots, L_{n}$ and $n=1, 2, \ldots, N$ by the signals received at the RRHs.

\subsection{Transmission Model}\label{ra}
The UE sends a signal $\sqrt{p_s}s(t)$,
in which $p_s$ is the transmitted energy, and $\mathbb{E}\{|s(t)|^2\!\}\!=\!1$. 
Given that the mmWave channel is sparse, we assume that $L_{n}+1\leq Q$, where $Q$ is the number of RF chains for each RRH.
The received signal $\mathbf{r}_{n}(t) \in \mathbb{C}^{Q\times 1}$ at RRH $n$ is given by \cite{c-channel}
\begin{equation}\label{receive}
\mathbf{r}_{n}(t)= \mathbf{A}\left(\sum_{l=0}^{L_{n}}\alpha_{n,l}^\circ \sqrt{p_s}s(t-\tau_{n,l}^\circ)\mathbf{a}(\phi_{n,l}^\circ,\theta_{n,l}^\circ)e^{j2\pi \nu_{n,l}^\circ t}\right)+\mathbf{n}(t),
\end{equation}
where
$\alpha_{n,l}^\circ$, $\tau_{n,l}^\circ$, $\phi_{n,l}^\circ$, $\theta_{n,l}^\circ$, and $\nu_{n,l}^\circ$ denote the complex gain, delay, azimuth AOA, elevation AOA, and Doppler shift for the $l$-th path, respectively;
$\mathbf{a}(\cdot)$ is the array response vector;
$\mathbf{A} \in \mathbb{C}^{Q\times K}$ is the combining matrix in the mmWave hybrid architecture; and $\mathbf{n}(t)  \in \mathbb{C}^{Q\times 1}$ is the zero-mean white Gaussian noise with a known power spectrum density.
The channel parameters ($\phi_{n,l}^\circ$, $\theta_{n,l}^\circ$, $\tau_{n,l}^\circ$, $\nu_{n,l}^\circ$), for $l=0, 1, \ldots, L_{n}$ and $n=1, 2, \ldots, N$ can be extracted from \eqref{receive} \cite{ce1,ce2,ce3}.
Here, $ (l = 0) $ represents the LOS path, and $ (l> 0) $ represents the NLOS path.
Localization can be embedded in either the initial access stage or data transmission stage without additional overhead.

\subsection{Relationship Between Channel and Location Parameters}\label{nf}
In this subsection, we map the channel parameters to the location parameters.

\begin{itemize}
\item \textbf{TDOA}:
For the LOS path, 
the distance between the UE and the RRH $n$ is 
\vspace{-0.25cm}
\begin{equation} \label{1}
r_n^\circ=v_c(\tau_{n,0}^\circ-\omega)=||\mathbf{u}^\circ-\mathbf{b}_n||,
\vspace{-0.25cm}
\end{equation}
where $v_c$ is the signal propagation speed, and $\omega$ is the unknown clock bias between CRAN and UE.
Without loss of generality, we define the TOA of LOS path received by the RRH $1$ $\tau_{1,0}^\circ$ as the reference time.
Then, the TDOA between the LOS path of the RRH $n$ and the reference time is
$\tau_{n,0}^\circ - \tau_{1,0}^\circ$.
Thus, we define the TDOA-related parameter as
\vspace{-0.25cm}
\begin{equation} \label{2}
r_{n1}^\circ= v_c(\tau_{n,0}^\circ - \tau_{1,0}^\circ) = r_n^\circ - r_1^\circ,
\vspace{-0.25cm}
\end{equation}
where the unknown $\omega$ can be eliminated.

For the NLOS path, we have
\vspace{-0.25cm}
\begin{equation} \label{sc0}
r_{n,l}^{s\circ} = v_c(\tau_{n,l}^s - \omega)= ||\mathbf{u}^\circ-\mathbf{s}_{n,l}^\circ|| + ||\mathbf{s}_{n,l}^\circ - \mathbf{b}_n||.
\vspace{-0.25cm}
\end{equation}
Then, the TDOA between the $l$-th NLOS path of the $n$-th RRH and the reference time is $\tau_{n,l}^\circ - \tau_{1,0}^\circ$,
and we define the TDOA-related parameter as
\vspace{-0.25cm}
\begin{equation} \label{sc1}
r_{n1,l}^{s\circ} = v_c(\tau_{n,l}^\circ - \tau_{1,0}^\circ) = r_{n,l}^{s\circ} - r_1^\circ.
\vspace{-0.25cm}
\end{equation}
Therefore, $r_{n1}^\circ$ and $r_{n1,l}^{s\circ}$ are the TDOA-related parameters, which are used in our proposed algorithms and are derived from the TDOA by multiplying with $v_c$.

\item \textbf{FDOA}:
For the LOS path, we define the time derivative of $r_n^\circ$ in \eqref{1} as $\dot{r}_{n}^\circ$, and we have
\begin{equation}\label{3}
\dot{r}_{n}^\circ= \dfrac{\partial {r}_{n}^\circ}{\partial t} =
\dfrac{\dot{\mathbf{u}}^{\circ T}\mathbf{u}^\circ+\mathbf{u}^{\circ T}\dot{\mathbf{u}}^\circ-2\dot{\mathbf{u}}^{\circ T}\mathbf{b}_n}{2\sqrt{(\mathbf{u}^\circ-\mathbf{b}_n)^T (\mathbf{u}^\circ-\mathbf{b}_n)}}=\dfrac{\dot{\mathbf{u}}^{\circ T}(\mathbf{u}^\circ-\mathbf{b}_n)}{||\mathbf{u}^\circ - \mathbf{b}_n||},
\end{equation}
which is the relative velocity between UE and RRH $n$.
Without loss of generality, we define the FOA or Doppler shift of LOS path received by the RRH $1$ $\nu_{1,0}^\circ$ as the reference frequency.
Then, the FDOA between the LOS path of the RRH $n$ and the reference frequency is
$\nu_{n,0}^\circ - \nu_{1,0}^\circ$.
Thus, given the signal wavelength $\lambda_c$, we obtain the FDOA-related parameter as
\vspace{-0.25cm}
\begin{equation} \label{4}
\dot{r}_{n1}^\circ= \lambda_c(\nu_{n,0}^\circ - \nu_{1,0}^\circ)=\dot{r}_n^\circ-\dot{r}_1^\circ.
\vspace{-0.25cm}
\end{equation}
For the NLOS path, 
we define the time derivative of $r_{n,l}^{s\circ}$ in \eqref{sc0} as $\dot{r}_{n,l}^{s\circ}$, and we have
\begin{equation} \label{sc4}
\dot{r}_{n,l}^{s\circ}=\dfrac{(\dot{\mathbf{u}}^\circ-\dot{\mathbf{s}}_{n,l}^\circ)^{ T}(\mathbf{u}^\circ-\mathbf{s}_{n,l}^\circ)}{||\mathbf{u}^\circ-\mathbf{s}_{n,l}^\circ||}+\dfrac{\dot{\mathbf{s}}_{n,l}^{\circ T}(\mathbf{s}_{n,l}^\circ - \mathbf{b}_n)}{||\mathbf{s}_{n,l}^\circ - \mathbf{b}_n||}.
\end{equation}
Then, the FDOA between the $l$-th NLOS path of the $n$-th RRH and the reference frequency is $\nu_{n,l}^\circ - \nu_{1,0}^\circ$,
and we obtain the FDOA-related parameter as
\vspace{-0.25cm}
\begin{equation} \label{44}
\dot{r}_{n1,l}^{s\circ} =\lambda_c(\nu_{n,l}^\circ - \nu_{1,0}^\circ) =\dot{r}_{n,l}^{s\circ} - \dot{r}_1^\circ.
\vspace{-0.25cm}
\end{equation}
Thus, $\dot{r}_{n1}^\circ$ and $\dot{r}_{n1,l}^{s\circ}$ are the FDOA-related parameters, which are used in our proposed algorithms, and they are derived from the FDOA by multiplying with $\lambda_c$.

\item \textbf{AOA}:
For the LOS path, we get  
\begin{equation} \label{5}
\phi_{n}^\circ=\phi_{n,0}^\circ= \arctan\frac{y^\circ-y_n^b}{x^\circ-x_n^b},  \ \ \ \ \ \theta_{n}^\circ=\theta_{n,0}^\circ=\arcsin\frac{z^\circ-z_n^b}{||\mathbf{u}^\circ-\mathbf{b}_n||}.
\end{equation}
Then, for the NLOS path, we have
\begin{equation} \label{55}
\phi_{n,l}^{s\circ} = \phi_{n,l}^\circ= \arctan\frac{y_{n,l}^{s\circ}-y_n^b}{x_{n,l}^{s\circ}-x_n^b} ,  \ \ \ \theta_{n,l}^{s\circ} = \theta_{n,l}^\circ=\arcsin\frac{z_{n,l}^{s\circ}-z_n^b}{||\mathbf{s}_{n,l}^\circ-\mathbf{b}_n||}.
\end{equation}
Thus, $(\phi_{n}^\circ,\theta_{n}^\circ)$ and $(\phi_{n,l}^{s\circ},\theta_{n,l}^{s\circ})$ are the AOA-related parameters.
\end{itemize}

Summarizing, the relationships between TDOA/FDOA/AOA-related channel and location parameters are given in \eqref{2}, \eqref{sc1}, \eqref{4}, \eqref{44}, \eqref{5}, and \eqref{55}.
In the following sections, we focus on developing effective algorithms to estimate the unknown location and velocity of the UE and scatterers as accurate as possible
by utilizing hybrid TDOA/FDOA/AOA measurements.
This task is not trivial given that the relations are nonlinear and nonconvex functions of $\mathbf{u}^\circ$, $\dot{\mathbf{u}}^\circ$, $\mathbf{s}_{n,l}^\circ$, and  $\dot{\mathbf{s}}_{n,l}^{\circ}$. 

\section{Problem Formulation}
\subsection{Measurement Selection}\label{da}
Each RRH obtains a set of measurements,
$\mathbb{M}_n = \{(\phi_{n,m}, \theta_{n,m}, \tau_{n,m}, \nu_{n,m})|m=1, 2, \ldots, M_{n}\}$ for $n=1, 2, \ldots, N$, where $M_n$ 
is the estimated number of paths, which is $L_n+1$ for a perfect value.
In this subsection, we propose a method to select measurements of $N_a$ LOS paths from all obtained $\sum_{n=1}^{N} M_n$ paths.

First, 
select the $m_n^*$-th path in $\mathbb{M}_n$, which has the smallest delay (probably the LOS path) among $M_{n}$ paths, for $n=1, 2, \ldots, N$.
Second, we define the rough estimate of $\mathbf{u}^{\circ}$ by the RRH $n$ to further eliminate NLOS paths in the chosen $N$ paths as follows
\vspace{-0.25cm}
\begin{equation}\label{a1}
\hat{\mathbf{u}}_n=\mathbf{b}_n+ v_c\tau_{n,m_n^*}[\cos\theta_{n,m_n^*}\cos\phi_{n,m_n^*},\  \cos\theta_{n,m_n^*}\sin\phi_{n,m_n^*},\ \sin\theta_{n,m_n^*}]^T,
\vspace{-0.25cm}
\end{equation}
for $n=1, 2, \ldots, N$.
Points in set $\{{\hat{\mathbf{u}}_n} | n=1, 2, \ldots, N\}$ 
are close and dispersed to one another if they are generated by LOS and NLOS measurements, respectively.
Subsequently, we classify $\hat{\mathbf{u}}_n$ for $n=1, 2, \ldots, N$ into two classes by K-means algorithm and obtain two class centers, namely, $\mathbf{c}_{\rm LOS}$ and $\mathbf{c}_{\rm NLOS}$. As the energy of LOS paths is much greater than that of NLOS paths in mmWave frequencies, we can further eliminate the NLOS paths in $\mathbf{c}_{\rm LOS}$.
A threshold is set according to the energy gap between the LOS and NLOS paths to determine the value of $N_a$. 
Then, the set of selected LOS measurements is $\mathbb{M}_{a}$.
The set of remaining measurements is $\mathbb{M}_{r,n}$, where $\mathbb{M}_{r,n}\cap\mathbb{M}_{a}=\emptyset$ for $n=1, 2, \ldots, N$.
	
We aim to estimate the unknown $\mathbf{u}^{\circ}$ and $\dot{\mathbf{u}}^{\circ}$ from the measurements in $\mathbb{M}_{a}$ (LOS measurements) and the unknown $\mathbf{s}_{n,l}^{\circ }$ and $\dot{\mathbf{s}}_{n,l}^{\circ}$ from the measurements in $\mathbb{M}_{r,n}$ (nearly all NLOS measurements) for $l=1, 2, \ldots, |\mathbb{M}_{r,n}|$ and $n=1, 2, \ldots, N$ as accurately as possible.

\subsection{Possible Solution}\label{BBNN}
\begin{figure*}
	\vspace{-0.5cm}
	\centering
	\includegraphics[scale=0.8,angle=0]{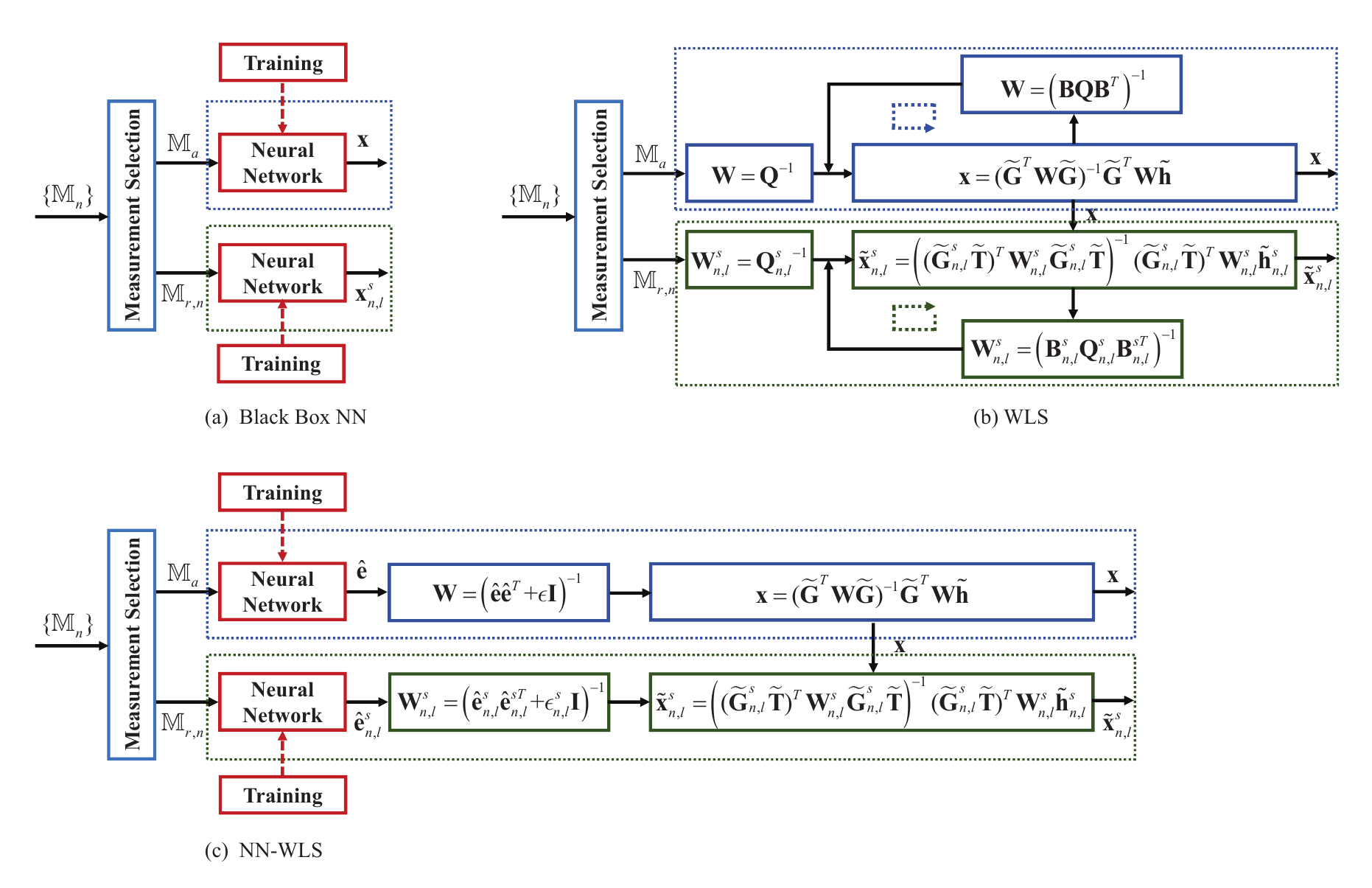}
	\captionsetup{font=footnotesize}
	\caption{Block diagrams of (a) Black Box NN; (b) WLS; (c) NN-WLS, where $\mathbf{x}=[\mathbf{u}^{T}, \dot{\mathbf{u}}^{T}]^T$ and  $\mathbf{x}_{n,l}^{s} = [\mathbf{s}_{n,l}^{T},\dot{\mathbf{s}}_{n,l}^{T}]^T$ denote the estimated location and velocity of UE and scatterer, respectively.}
	\label{NN}
\end{figure*}

After the measurement selection process, the corresponding measurements in sets $\mathbb{M}_{a}$ and $\mathbb{M}_{r,n}$ can be fed into black box NNs and trained end-to-end using real datasets to learn $\mathbf{x}^\circ=[\mathbf{u}^{\circ T}, \dot{\mathbf{u}}^{\circ T}]^T$ and  $\mathbf{x}_{n,l}^{s\circ} = [\mathbf{s}_{n,l}^{\circ T},\dot{\mathbf{s}}_{n,l}^{\circ T}]^T$ directly, respectively (Fig. \ref{NN}(a)).
However, the localization accuracy of this method is limited, and a prohibitively large amount of training data is required to improve the localization accuracy.
To skip this step, 
our strategy is based on the argument that the model is mathematically well developed with fewer uncertainties \cite{md}.
However, the model generally relies on some approximations and ideal assumptions, which worsen the performance when the measurement noise increases.
Motivated by the powerful learning ability of the NN, its use to replace the approximate operations in the model can further improve the performance.
Therefore, we combine NNs with geometric models in this study.
Specifically,
we first develop an unbiased model-based WLS localization estimator (Section \ref{wls1}).
Then, we establish a NN-assisted WLS localization method (Section \ref{NNWLS}) by introducing NNs into the developed WLS model (or estimator)
to learn the higher-order error components, 
thereby improving the performance of the estimator, especially in a large noisy environment.

\section{Model-based WLS Localization}\label{wls1}
In this section, we devise a closed-form localization
estimator that approximates the maximum likelihood (ML) estimator
under small noise conditions.
We improve the performance of the traditional multi-stage WLS estimator \cite{l3} by exploiting angular information and establishing a one-stage WLS estimator. We further extend the traditional WLS localization estimator that can only be used for UE localization to the scatterers' localization.

Measurements in $\mathbb{M}_{a}$ are used to estimate the 
location and velocity of the UE.
According to \eqref{2},  \eqref{4}, and \eqref{5},
we denote a noise-free vector of hybrid TDOA/FDOA/AOA parameters as
$\mathbf{m}^\circ=[r_{21}^\circ, \dot{r}_{21}^\circ, \ldots, r_{{N_{a}}1}^\circ, \dot{r}_{{N_{a}}1}^\circ,\phi_1^\circ, \theta_1^\circ, \ldots,\phi_{N_{a}}^\circ, \theta_{N_{a}}^\circ]^T$.
Then, we model the hybrid measurements by the additive noise model as 
$\mathbf{m}\!=\!\mathbf{m}^\circ\!+\!\Delta \mathbf{m}$,
where 
$\mathbf{m}\!=\![r_{21}, \dot{r}_{21},\! \ldots\!, r_{{N_{a}}1}, \dot{r}_{{N_{a}}1},\phi_1, \theta_1,\!\ldots\!,\phi_{N_{a}}, \theta_{N_{a}}]^T\!\!,$
and $\Delta \mathbf{m}=[\Delta r_{21}, \Delta\dot{r}_{21}, \ldots, \Delta r_{N_{a}1}, \Delta\dot{r}_{N_{a}1}, \Delta\phi_1, \Delta\theta_1, \ldots, \Delta\phi_{N_{a}}, \Delta\theta_{N_{a}}]^T$ is a Gaussian noise vector with zero mean and covariance matrix $\mathbf{Q}$.
Measurements in $\mathbb{M}_{r,n}$ are used to estimate the location and velocity of scatterers for $n=1, 2, \ldots, N$.
According to \eqref{sc1},  \eqref{44}, and \eqref{55},
we obtain the $l$-th noise-free vector of hybrid parameters as $\mathbf{m}_{n,l}^{s\circ}=[r_{n1,l}^{s\circ},\dot{r}_{n1,l}^{s\circ}, \phi_{n,l}^{s\circ}, \theta_{n,l}^{s\circ}]^T$,
and hybrid measurements 
$\mathbf{m}_{n,l}^{s} = \mathbf{m}_{n,l}^{s\circ} + \Delta\mathbf{m}_{n,l}^{s},$
where 
$\mathbf{m}_{n,l}^{s}=[r_{n1,l}^{s},\dot{r}_{n1,l}^{s}, \phi_{n,l}^{s}, \theta_{n,l}^{s}]^T$,
and $\Delta\mathbf{m}_{n,l}^{s}=[\Delta r_{n1,l}^{s},\Delta \dot{r}_{n1,l}^{s}, \Delta\phi_{n,l}^{s}, \Delta\theta_{n,l}^{s}]^T$ with  zero mean and covariance matrix $\mathbf{Q}^s_{n,l}$,
for $l=1, 2, \ldots, |\mathbb{M}_{r,n}|$.

\subsection{UE Localization}\label{ps}
In this subsection, we present a closed-form method for estimating
the UE location and velocity.
We first establish a set of pseudo-linear TDOA and FDOA
equations by nonlinear transformation and AOA exploitation.
Subsequently, AOA equations are derived and combined with
TDOA and FDOA equations to obtain an accurate estimation.

First, we derive $2(N_{a}-1)$ pseudo-linear TDOA and FDOA equations.
We rewrite \eqref{2} as $r_{n1}^\circ+r_1^\circ=r_{n}^\circ$ and square both sides to yield $(r_{n1}^\circ)^2+2r_{n1}^\circ r_1^\circ=(r_{n}^\circ)^2-(r_1^\circ)^2$.
According to \eqref{1}, we obtain
\vspace{-0.25cm}
\begin{equation} \label{8}
(r_{n1}^\circ)^2+2r_{n1}^\circ r_1^\circ=\mathbf{b}_n^T\mathbf{b}_n-\mathbf{b}_1^T\mathbf{b}_1-2(\mathbf{b}_n-\mathbf{b}_1)^T \mathbf{u}^\circ.
\vspace{-0.25cm}
\end{equation}
Equation \eqref{8} is pseudo-linear formula with respect to $\mathbf{u}^\circ$ and $r_1^\circ$.
Then, by taking the time derivative of \eqref{8}, we yield
\vspace{-0.25cm}
\begin{equation} \label{12}
\dot{r}_{n1}^{\circ}{r}_{n1}^{\circ}+\dot{r}_{n1}^{\circ}r_1^\circ+r_{n1}^{\circ}\dot{r}_1^\circ=(\mathbf{b}_1-\mathbf{b}_n)^T \dot{\mathbf{u}}^\circ.
\vspace{-0.25cm}
\end{equation}
Equation \eqref{12} is pseudo-linear formula with respect to $\dot{\mathbf{u}}^\circ$, $r_1^\circ$, and $\dot{r}_1^\circ$.
However, $r_1^\circ$ and $\dot{r}_1^\circ$ cannot be obtained directly from the channel measurements TDOA and FDOA.
A well-known solution for localization that uses TDOAs and FDOAs is to
find $\mathbf{u}^\circ$ and $\dot{\mathbf{u}}^\circ$ by using multi-stage WLS estimators \cite{l3}.
The conventional method is based on the estimation of the
redundant parameters, namely, $r_1^\circ$ and $\dot{r}_1^\circ$, together with the UE location and velocity. 
In the present study, we apply a different approach, that is, we use AOA measurements to eliminate the redundant parameters to estimate $\mathbf{u}^\circ$ and $\dot{\mathbf{u}}^\circ$ in only one stage.

To eliminate $r_1^\circ$ and $\dot{r}_1^\circ$ in \eqref{8} and \eqref{12},
we define $\mathbf{a}_1^\circ=[\cos\theta_1^\circ\cos\phi_1^\circ, \cos\theta_1^\circ\sin\phi_1^\circ, \sin\theta_1^\circ]^T$, which is a unit-norm angular vector that possesses the properties:
$\mathbf{a}_1^{\circ T}\mathbf{a}_1^\circ=1$ and $\dot{\mathbf{a}}_1^{\circ T}\mathbf{a}_1^\circ=\mathbf{a}_1^{\circ T}\dot{\mathbf{a}}_1^\circ=0$.
Multiplying both sides of \eqref{8} by $\mathbf{a}_1^{\circ T}\mathbf{a}_1^\circ$ 
and utilizing the geometric relationship
$\mathbf{u}^\circ-\mathbf{b}_1=r_1^\circ \mathbf{a}_1^\circ$
yield
\vspace{-0.25cm}
\begin{equation} \label{11}
(r_{n1}^\circ)^2-2r_{n1}^\circ \mathbf{a}_1^{\circ T}\mathbf{b}_1-\mathbf{b}_n^T \mathbf{b}_n+\mathbf{b}_1^T \mathbf{b}_1=2[(\mathbf{b}_1-\mathbf{b}_n)^T-r_{n1}^\circ \mathbf{a}_1^{\circ T}]\mathbf{u}^\circ.
\vspace{-0.25cm}
\end{equation}
Multiplying both sides of \eqref{12} by $\mathbf{a}_1^{\circ T}\mathbf{a}_1^\circ$ 
and utilizing the geometric relationship $\dot{\mathbf{u}}^\circ=\dot{r}_1^\circ \mathbf{a}_1^\circ+r_1^\circ\dot{\mathbf{a}}_1^\circ$
yield
\vspace{-0.25cm}
\begin{equation} \label{16}
\dot{r}_{n1}^{\circ}{r}_{n1}^{\circ}-\dot{r}_{n1}^{\circ}\mathbf{a}_1^{\circ T}\mathbf{b}_1=-\dot{r}_{n1}^{\circ}\mathbf{a}_1^{\circ T}\mathbf{u}^\circ+[(\mathbf{b}_1-\mathbf{b}_n)^T-{r}_{n1}^{\circ}\mathbf{a}_1^{\circ T}]\dot{\mathbf{u}}^\circ.
\vspace{-0.25cm}
\end{equation}
By collecting (\ref{11}) and (\ref{16}), for $n=2,\ldots, N_{a}$, $2(N_{a}-1)$ pseudo-linear TDOA and FDOA equations are obtained.
Then, we derive $2N_{a}$ AOA equations for $n=1,2,\ldots, N_{a}$ according to \eqref{5}, which is given by
\vspace{-0.25cm}
\begin{equation} \label{17}
\mathbf{c}_n^{\circ T}\mathbf{b}_n=\mathbf{c}_n^{\circ T}\mathbf{u}^\circ, \ \ \mathbf{d}_n^{\circ T}\mathbf{b}_n=\mathbf{d}_n^{\circ T}\mathbf{u}^\circ,
\vspace{-0.25cm}
\end{equation}
where $\mathbf{c}_n^\circ=[-\sin\phi_n^\circ, \cos\phi_n^\circ, 0]^T$ and $\mathbf{d}_n^\circ=[-\sin\theta_n^\circ\cos\phi_n^\circ, -\sin\theta_n^\circ\sin\phi_n^\circ, \cos\theta_n^\circ]^T$.
Collecting \eqref{11}, \eqref{16} for $n=2,\ldots, N_{a}$ and  \eqref{17} for $n=1,2,\ldots, N_{a}$ yields the following matrix equation,
\vspace{-0.25cm}
\begin{equation} \label{191}
\mathbf{h}=\mathbf{G}\mathbf{x}^\circ,
\vspace{-0.25cm}
\end{equation}
where $\mathbf{x}^\circ=[\mathbf{u}^{\circ T}, \dot{\mathbf{u}}^{\circ T}]^T$ is an unknown six-dimensional vector of location and velocity of the UE, and 
\vspace{-0.5cm}
\begin{equation}
\begin{split}
&\mathbf{h}=[\mathbf{q}_2^T,\dots,\mathbf{q}_{N_{a}}^T, \mathbf{h}_1^T, \dots,\mathbf{h}_{N_{a}}^T]^T,\ \ \mathbf{G}=[\mathbf{P}_2^T,\dots,\mathbf{P}_{N_{a}}^T, \mathbf{G}_1^T, \dots,\mathbf{G}_{N_{a}}^T]^T,
\end{split}
\end{equation}
\begin{equation}
\begin{split}
&\mathbf{q}_n\!=\!\begin{pmatrix}
(r_{n1}^\circ)^2\!-\!2r_{n1}^\circ \mathbf{a}_1^{\circ T}\mathbf{b}_1\!-\!\mathbf{b}_n^T \mathbf{b}_n\!+\!\mathbf{b}_1^T \mathbf{b}_1 \\
\dot{r}_{n1}^{\circ}{r}_{n1}^{\circ}-\dot{r}_{n1}^{\circ}\mathbf{a}_1^{\circ T}\mathbf{b}_1\\
\end{pmatrix},\ \ \ \ \ \ \ \ \ \ \ \ \ \mathbf{h}_n\!=\!\begin{pmatrix}
\mathbf{c}_n^{\circ T}\!\mathbf{b}_n \\
\mathbf{d}_n^{\circ T}\!\mathbf{b}_n
\end{pmatrix},\\
\end{split}
\end{equation}
\begin{equation}
\begin{split}
&\mathbf{P}_n\!=\!\begin{pmatrix}\!
2[(\mathbf{b}_1\!-\!\mathbf{b}_n)^{T}\!-\!r_{n1}^{\circ}\mathbf{a}_1^{\circ T}] & \mathbf{0}^T\\
-\dot{r}_{n1}^{\circ}\mathbf{a}_1^{\circ T} & (\!\mathbf{b}_1\!-\!\mathbf{b}_n\!)^{T}\!-\!r_{n1}^{\circ}\mathbf{a}_1^{\circ T}\!
\end{pmatrix}\!, \ \  \mathbf{G}_n\!=\!\begin{pmatrix}
\mathbf{c}_n^{\circ T} \!&\! \mathbf{0}^T\\
\mathbf{d}_n^{\circ T} \!&\! \mathbf{0}^T
\end{pmatrix}\!,
\vspace{-0.25cm}
\end{split}
\end{equation}
where $\mathbf{0}$ is a $3\times 1$ zero vector.
Equation \eqref{191} is the noise-free matrix representation of the joint location and velocity estimation model.

The noise-free parameters in vector $\mathbf{h}$ and matrix $\mathbf{G}$ in (\ref{191}) are not available.
Let the noisy measurements replace the noise-free parameters in $\mathbf{h}$ and $\mathbf{G}$ (i.e., let $r_{i1}\!=\!r_{i1}^{\circ}\!+\!\Delta r_{i1}$, $\dot{r}_{i1}=\dot{r}_{i1}^{\circ}+\Delta \dot{r}_{i1}$, $\phi_j=\phi_j^{\circ}+\Delta \phi_j$, and
$\theta_j=\theta_j^{\circ}+\Delta\theta_j$ replace $r_{i1}^{\circ}$, $\dot{r}_{i1}^{\circ}$, $\phi_j^{\circ}$, and $\theta_j^{\circ}$, for $i=2,\ldots,N_{a}$ and $j=1,\ldots,N_{a}$), we define the error vector
\vspace{-0.25cm}
\begin{equation}\label{20}
\mathbf{e}=\tilde{\mathbf{h}}-\tilde{\mathbf{G}}\mathbf{x}^{\circ},
\vspace{-0.25cm}
\end{equation}
where $\tilde{\mathbf{h}}$ and $\tilde{\mathbf{G}}$ are the noisy counterparts.
The WLS solution \cite{KAY} of $\mathbf{x}^{\circ}$ 
can be obtained as
\vspace{-0.25cm}
\begin{equation}\label{21}
\mathbf{x}=(\tilde{\mathbf{G}}^{T}\mathbf{W}\tilde{\mathbf{G}})^{-1}\tilde{\mathbf{G}}^{T}\mathbf{W}\tilde{\mathbf{h}},
\vspace{-0.25cm}
\end{equation}
where the weighting matrix $\mathbf{W}=(\mathbb{E}\{\mathbf{e}\mathbf{e}^{T}\})^{-1}$.
In view of the nonlinearity of $\mathbf{e}$, obtaining the weighting matrix $\mathbf{W}$ is difficult in general. 
By ignoring the second- and higher-order noise terms, we approximate $\mathbf{e}$ with its linear terms as 
\vspace{-0.45cm}
\begin{equation}\label{e}
\mathbf{e} \approx \mathbf{B}\Delta \mathbf{m},
\vspace{-0.7cm}
\end{equation}
where
\begin{eqnarray}\label{b}
&&\mathbf{B}=\begin{bmatrix}
\mathbf{B}_1&\mathbf{B}_2\\
\mathbf{O}&\mathbf{B}_3
\end{bmatrix}, \ \ \mathbf{B}_1=\mbox{blkdiag}\left(\begin{bmatrix} 2r_2^{\circ}&0\\ \dot{r}_2^{\circ}&r_2^{\circ} \end{bmatrix}, \ldots,  \begin{bmatrix} 2r_{N_a}^{\circ}&0\\ \nonumber
\dot{r}_{N_a}^{\circ}&r_{N_a}^{\circ}
\end{bmatrix}\right),\\
&&\mathbf{B}_2=\begin{bmatrix}\mathbf{B}_{21}&\mathbf{O}
\end{bmatrix},\ \  
\mathbf{B}_{21}=\begin{bmatrix}
0&0;\ a_2&b_2;\ \ldots;\ 0&0;\ a_{N_a}&b_{N_a}
\end{bmatrix}, \\ \nonumber
&&\mathbf{B}_3=\mbox{diag}\left(
r_1^{\circ}\cos\theta_1^{\circ},r_1^{\circ},\ldots, r_{N_a}^{\circ}\cos\theta_{N_a}^{\circ},r_{N_a}^{\circ} \right),
\end{eqnarray}
in which the ``;" operator separates the rows in a matrix;
$a_n = r_1^{\circ}r_{n1}^{\circ}\dot{\phi}_1^{\circ}\cos^2\theta_1^{\circ}$ and $b_n = r_1^{\circ}r_{n1}^{\circ}\dot{\theta}_1^{\circ}$ for $n=2,\ldots,N_{a}$;
$\dot{\phi}_1^{\circ}= {\mathbf{c}_1^{\circ T}\dot{\mathbf{u}}^\circ}/({r_1^\circ\cos\theta_1^\circ})$ and $\dot{\theta}_1^{\circ}= {\dot{\mathbf{u}}^{\circ T}\mathbf{d}_1^\circ}/{r_1^\circ}$ are the time derivatives of
\eqref{5} with $n=1$.
The detailed derivations of \eqref{e} are listed in Appendix \ref{B}.
As we approximate $\mathbf{e}$ up to its linear noise term $\mathbf{B}\Delta \mathbf{m}$, 
it follows from the distribution of $\Delta \mathbf{m}$ that $\mathbf{e}$ is a zero-mean Gaussian vector with covariance matrix $\mathbf{B}\mathbf{Q}\mathbf{B}^T$.
Therefore, the weighting matrix can be easily calculated as
\vspace{-0.25cm}
\begin{equation}\label{w}
\mathbf{W}=\left(\mathbf{B}\mathbf{Q}\mathbf{B}^T\right)^{-1},
\vspace{-0.25cm}
\end{equation}
where the weighting matrix $\mathbf{W}$ is dependent on the unknown location $\mathbf{u}^\circ$ and velocity $\dot{\mathbf{u}}^\circ$ via the matrix $\mathbf{B}$.
Hence, we initialize $\mathbf{W} =\mathbf{Q}^{-1}$ to
provide the initial location and velocity estimates.
Updating this initial solution in $\mathbf{B}$ can construct a more accurate weighting matrix by \eqref{w} to derive the final solutions of $\mathbf{u}^{\circ}$ and $\dot{\mathbf{u}}^{\circ}$.

\vspace{-0.25cm}
\subsection{Scatterer Localization}
In this subsection, we present a closed-form method for estimating
the scatterers' location and velocity.
We take the $l$-th scatterer between the $n$-th RRH and the UE 
for example, where $1 \leqslant l \leqslant |\mathbb{M}_{r,n}|$.
First, let $d_{1,n,l}^{\circ} = ||\mathbf{s}_{n,l}^\circ-\mathbf{b}_n||$, $d_{2,n,l}^\circ = ||\mathbf{u}^\circ-\mathbf{s}_{n,l}^\circ||$, and we have
$r_{n,l}^{s\circ}=d_{1,n,l}^{\circ} + d_{2,n,l}^\circ$.
By rewriting \eqref{sc1} as
$r_{n1,l}^{s\circ}+r_1^{\circ}-d_{1,n,l}^\circ = d_{2,n,l}^\circ$, squaring both sides, 
and making some simplifications,
we obtain
\vspace{-0.5cm}
\begin{equation} \label{s4}
(r_{n1,l}^{s\circ}+r_1^{\circ})^2 - 2 (r_{n1,l}^{s\circ}+r_1^{\circ}) d_{1,n,l}^\circ ={\mathbf{u}^\circ}^T\mathbf{u}^\circ-2{\mathbf{u}^\circ}^T\mathbf{s}_{n,l}^\circ + 2 {\mathbf{b}_n^T}\mathbf{s}_{n,l}^\circ - {\mathbf{b}_n^T}\mathbf{b}_n.
\vspace{-0.25cm}
\end{equation}
Then, by taking the time derivative of \eqref{s4}, we have
\vspace{-0.25cm}
\begin{equation} \label{s44}
(r_{n1,l}^{s\circ}+r_1^{\circ})(\dot{r}_{n1,l}^{s\circ}+\dot{r}_1^{\circ})  -  (\dot{r}_{n1,l}^{s\circ}+\dot{r}_1^{\circ}) d_{1,n,l}^\circ - (r_{n1,l}^{s\circ}+r_1^{\circ}) \dot{d}_{1,n,l}^\circ ={\dot{\mathbf{u}}^{\circ T}} \mathbf{u}^\circ
-{\dot{\mathbf{u}}^{\circ T}}\mathbf{s}_{n,l}^\circ 
-{\mathbf{u}^{\circ T}}\dot{\mathbf{s}}_{n,l}^\circ 
+ {\mathbf{b}_n^T}\dot{\mathbf{s}}_{n,l}^\circ,
\vspace{-0.25cm}
\end{equation}
where $\dot{d}_{1,n,l}^\circ$ is the time derivation of ${d}_{1,n,l}^\circ$.
By utilizing the AOA parameters, together with estimated $\mathbf{u}^{\circ}$ and $\dot{\mathbf{u}}^{\circ}$ in Section \ref{ps}, we can eliminate the redundant parameters (${d}_{1,n,l}^\circ$ and $\dot{d}_{1,n,l}^\circ$)
in \eqref{s4} and \eqref{s44} to estimate $\mathbf{s}_{n,l}^\circ$ and $\dot{\mathbf{s}}_{n,l}^\circ$ in one stage.
${r}_1^{\circ}$ and $\dot{r}_1^{\circ}$ are obtained by estimated $\mathbf{u}^{\circ}$ and $\dot{\mathbf{u}}^{\circ}$, thus, $r_{n,l}^{s\circ}=r_{n1,l}^{s\circ}+r_1^{\circ}$ and $\dot{r}_{n,l}^{s\circ}\!=\!\dot{r}_{n1,l}^{s\circ}+\dot{r}_1^{\circ}$ are obtained.
By defining $\mathbf{a}_{n,l}^{s\circ} \! =\! [\cos\theta_{n,l}^{s\circ}\cos\phi_{n,l}^{s\circ}, \cos\theta_{n,l}^{s\circ}\sin\phi_{n,l}^{s\circ}, \sin\theta_{n,l}^{s\circ}]^T\!$, 
eliminating ${d}_{1,n,l}^\circ$ and $\dot{d}_{1,n,l}^\circ$ in \eqref{s4} and \eqref{s44}, 
and combining AOA equations,
we obtain the following matrix representation,
\vspace{-0.25cm}
\begin{equation} \label{s12}
\mathbf{h}^{s}_{n,l}=\mathbf{G}^{s}_{n,l}\mathbf{x}_{n,l}^{s\circ},
\vspace{-0.25cm}
\end{equation}
where 
\vspace{-0.25cm}
\begin{eqnarray*} \label{s13}
\mathbf{h}^{s}_{n,l}\!\! =\!\! \begin{pmatrix}\!
(r_{n,l}^{s\circ})^2 \!+\! 2 r_{n,l}^{s\circ}{\mathbf{a}_{n,l}^{s\circ}}^T\mathbf{b}_n \!-\! {\mathbf{u}^\circ}^T\mathbf{u}^\circ \!+\! {\mathbf{b}_n^T}\mathbf{b}_n\\
r_{n,l}^{s\circ}\dot{r}_{n,l}^{s\circ} + \dot{r}_{n,l}^{s\circ}{\mathbf{a}_{n,l}^{s\circ}}^T\mathbf{b}_n - {\dot{\mathbf{u}}^{\circ T}}\mathbf{u}^\circ\\
{\mathbf{c}_{n,l}^{s\circ}}^T\mathbf{b}_n\\
{\mathbf{d}_{n,l}^{s\circ}}^T\mathbf{b}_n
\!\!\end{pmatrix}\!,\ \mathbf{G}^{s}_{n,l} \!\!=\!\! \begin{pmatrix}\!
2({\mathbf{b}_n}\!-\!{\mathbf{u}^\circ}\!+\!r_{n,l}^{s\circ} {\mathbf{a}_{n,l}^{s\circ}})^T\!\!\!&\!\!\mathbf{0}^T\\
(\dot{r}_{n,l}^{s\circ} {\mathbf{a}_{n,l}^{s\circ}}-{\dot{\mathbf{u}}^\circ} )^T\!\!\!&\!\!({r}_{n,l}^{s\circ} {\mathbf{a}_{n,l}^{s\circ}} \!+\! \mathbf{b}_n \!-\!{{\mathbf{u}}^\circ} )^T \\
{\mathbf{c}_{n,l}^{s\circ}}^T\!\!\!&\!\!\mathbf{0}^T\\
{\mathbf{d}_{n,l}^{s\circ}}^T\!\!\!&\!\!\mathbf{0}^T
\!\!\end{pmatrix}\!,
\end{eqnarray*}
$\mathbf{c}_{n,l}^{s\circ}=[-\sin\phi_{n,l}^{s\circ}, \cos\phi_{n,l}^{s\circ}, 0]^T$, $\mathbf{d}_{n,l}^{s\circ}=[-\sin\theta_{n,l}^{s\circ}\cos\phi_{n,l}^{s\circ}, -\sin\theta_{n,l}^{s\circ}\sin\phi_{n,l}^{s\circ}, \cos\theta_{n,l}^{s\circ}]^T$,
and $\mathbf{x}_{n,l}^{s\circ} = [\mathbf{s}_{n,l}^{\circ T},\dot{\mathbf{s}}_{n,l}^{\circ T}]^T$.
However, four measurements are not enough for six unknowns.
We assume that the moving scatterers are vehicles that move along the same road as the UE and we can regard the road as straight within a short distance.
Thus, the direction of the scatterer velocity is aligned with the UE within a short distance. 
Let a unit vector $\mathbf{n}_{v}=\dot{\mathbf{u}}^{\circ}/\lVert\dot{\mathbf{u}}^{\circ}\lVert$ denote the direction of the UE velocity.
When $\dot{\mathbf{u}}^{\circ}$ is estimated in Section \ref{ps}, $\mathbf{n}_{v}$ is obtained.
We have $\dot{\mathbf{s}}_{n,l}^{\circ}=\dot{s}_{n,l}^{\circ}\mathbf{n}_{v}$, where $\dot{s}_{n,l}^{\circ}$ represents the magnitude of velocity.
With a transformation matrix,
\begin{equation}\label{t}
\mathbf{T}=\begin{pmatrix}
\mathbf{I}_{3\times 3}&\mathbf{0}\\
\mathbf{O}_{3\times 3}&\mathbf{n}_{v}
\end{pmatrix},
\vspace{-0.4cm}
\end{equation}
we obtain
\vspace{-0.1cm}
\begin{equation} \label{s122}
\mathbf{h}^{s}_{n,l}=\mathbf{G}^{s}_{n,l}\mathbf{T}\tilde{\mathbf{x}}_{n,l}^{s\circ},
\vspace{-0.25cm}
\end{equation}
where 
$\tilde{\mathbf{x}}_{n,l}^{s\circ} = [\mathbf{s}_{n,l}^{\circ T},\dot{s}_{n,l}^{\circ}]^T$ is an unknown four-dimensional vector of location and velocity magnitude of the scatterer.
\footnote{
		A scatterer can be in the opposite direction of a UE, because $\dot{s}_{n,l}^{\circ}$ can be negative.
		Moreover, we can judge whether the assumption that the scatterer is on the same road as the UE is met by comparing the estimated scatterer location with a rough offline map.
		If the assumption is satisfied, then we will believe the corresponding velocity estimate.
}
Replacing the noise-free parameters $\{r_{n1,l}^{s\circ},\dot{r}_{n1,l}^{s\circ}, \phi_{n,l}^{s\circ}, \theta_{n,l}^{s\circ}, \mathbf{u}^{\circ},\dot{\mathbf{u}}^{\circ}\}$ in \eqref{s122} by the noisy measurements $\{r_{n1,l}^{s},\dot{r}_{n1,l}^{s}, \phi_{n,l}^{s},  \theta_{n,l}^{s}\}$ and estimated $\{\mathbf{u},\dot{\mathbf{u}}\}$ results in the error vector
\vspace{-0.25cm}
\begin{equation} \label{s14}
\mathbf{e}_{n,l}^s = \tilde{\mathbf{h}}^{s}_{n,l} - \tilde{\mathbf{G}}^{s}_{n,l}\tilde{\mathbf{T}}\tilde{\mathbf{x}}_{n,l}^{s\circ}.
\vspace{-0.25cm}
\end{equation}
By approximating $\mathbf{e}_{n,l}^s$ up to the linear noise term, we have
$\mathbf{e}_{n,l}^s \approx \mathbf{B}_{n,l}^s\Delta \mathbf{m}_{n,l}^s,$
where
\vspace{-0.25cm}
\begin{equation}\label{s16}
\mathbf{B}_{n,l}^s=\begin{pmatrix}
2d_{2,n,l}^\circ&0&0&0\\
\dot{d}_{2,n,l}^\circ&d_{2,n,l}^\circ& -r_{n,l}^{s\circ}d_{1,n,l}^\circ\dot{\phi}_{n,l}^{s\circ}\cos^2\phi_{n,l}^{s\circ}&-r_{n,l}^{s\circ}d_{1,n,l}^\circ\dot{\theta}_{n,l}^{s\circ}\\ 
0&0&d_{n1}^\circ\cos\theta_n^{s\circ}&0\\
0&0&0&d_{n1}^\circ \end{pmatrix},
\vspace{-0.25cm}
\end{equation}
and 
$\dot{\phi}_{n,l}^{s\circ}= {\mathbf{c}_{n,l}^{s\circ T}\dot{\mathbf{s}}_{n,l}^\circ}/({d_{1,n,l}^\circ\cos\theta_{n,l}^{s\circ}})$, $\dot{\theta}_{n,l}^{s\circ}= {\dot{\mathbf{s}}_{n,l}^{\circ T}\mathbf{d}_1^\circ}/{d_{1,n,l}^\circ}$.
The derivations of \eqref{s16} are similar to those in Appendix \ref{B}, and we omit these details because of lack of space in this paper.
Thus, the WLS solution of $\tilde{\mathbf{x}}_{n,l}^{s\circ}$ is given by
\vspace{-0.3cm}
\begin{equation}\label{s17}
\tilde{\mathbf{x}}_{n,l}^{s}=
\left((\tilde{\mathbf{G}}^{s}_{n,l}\tilde{\mathbf{T}})^{T}\mathbf{W}^s_{n,l}\tilde{\mathbf{G}}^{s}_{n,l}\tilde{\mathbf{T}}\right)^{-1}
(\tilde{\mathbf{G}}^{s}_{n,l}\tilde{\mathbf{T}})^{T}
\mathbf{W}^s_{n,l}
\mathbf{\tilde{h}}^s_{n,l},
\vspace{-0.25cm}
\end{equation}
where $\mathbf{W}^s_{n,l}=\left(\mathbf{B}^s_{n,l}\mathbf{Q}^s_{n,l}{\mathbf{B}^{sT}_{n,l}}\right)^{-1}$.
The weighting matrix $\mathbf{W}^s_{n,l}$ is dependent on
$\mathbf{s}_{n,l}^{\circ}$ and $\dot{\mathbf{s}}_{n,l}^{\circ}$ 
through $\mathbf{B}^s_{n,l}$. 
At the beginning, we can use $\mathbf{W}^s_{n,l} ={(\mathbf{Q}^s_{n,l})}^{-1}$ in \eqref{s17} to produce a solution from which to generate a better $\mathbf{W}^s_{n,l}$  to yield a more accurate solution.

\vspace{-0.25cm}
\subsection{Discussion}
The proposed model-based localization method is summarized in Algorithm \ref{alg1} and Fig. \ref{NN}(b).
Repeating the solution computation one to two times in Algorithm \ref{alg1} (b) and (c) is sufficient to yield an accurate solution that reaches the CRLB for small Gaussian noise.

\begin{algorithm}	
	\caption{\textbf{: Pseudocode of the Proposed Model-based Localization Method}}\label{alg1}	
	\textbf{(a) Measurement Selection (Separate LOS and NLOS Measurements):}
	\begin{algorithmic}[1]	
		\Require $\mathbb{M}_{n}$ for $n=1, 2, \ldots, N$.
		\Ensure 	$\mathbb{M}_{a}$ and $\mathbb{M}_{r,n}$	for $n=1, 2, \ldots, N$.	
		\State Choose the $m_n^*$-th path in $\mathbb{M}_n$ which has the smallest delay among $M_{n}$ paths, for $n=1, 2, \ldots, N$.
		\State Calculate $\hat{\mathbf{u}}_n$ for $n=1, 2, \ldots, N$, according to \eqref{a1}. Classify $\hat{\mathbf{u}}_n$ for $n=1, 2, \ldots, N$ into two classes by K-means algorithm,
		and obtain two class centers $\mathbf{c}_{\rm LOS}$ and $\mathbf{c}_{\rm NLOS}$.		
		\State Calculate distance $d_n = ||\mathbf{c}_{\rm LOS}-\hat{\mathbf{u}}_n||$,
		and sort $d_n$ for $n=1, 2, \ldots, N$ in ascending order.
		Choose measurements of $N_a$ paths corresponding to the first $N_a$ smallest distances,
		and the set of selected measurements is $\mathbb{M}_{a}$.
		The set of remaining measurements is $\mathbb{M}_{r,n}$, for $n=1, 2, \ldots, N$.
	\end{algorithmic}

	\textbf{(b) UE Localization (Use LOS Measurements):}
	\begin{algorithmic}[1]				
		\Require $\mathbb{M}_{a}$. 
		\Ensure $\mathbf{x}=[\mathbf{u}^{ T}, \dot{\mathbf{u}}^{T}]^T$.
		\State Find $\mathbf{x}$ from (\ref{21}) with $\mathbf{W}=\mathbf{Q}^{-1}$.	
		\Repeat
		\State Calculate the matrix $\mathbf{B}$ in \eqref{b} by the obtained $\mathbf{x}$.
		\State Update the weighting matrix $\mathbf{W}$ in \eqref{w} by the obtained $\mathbf{B}$.
		\State Find $\mathbf{x}$ from (\ref{21}) with new  $\mathbf{W}$.
		\Until{convergence}
	\end{algorithmic}

\textbf{(c) Scatterer Localization (Use NLOS Measurements):}
\begin{algorithmic}[1]
	\Require $\mathbf{u}$, $\dot{\mathbf{u}}$, and $\mathbb{M}_{r,n}$, for $n=1, 2, \ldots, N$. 
	\Ensure $\tilde{\mathbf{x}}_{n,l}^{s} = [\mathbf{s}_{n,l}^{T},\dot{s}_{n,l}]^T$, for $l=1, 2, \ldots, |\mathbb{M}_{r,n}|$ and $n=1, 2, \ldots, N$. 	
	\For {$n=1$ to $N$}
	\For  {$l=1$ to $|\mathbb{M}_{r,n}|$}
    \State Find $\tilde{\mathbf{x}}_{n,l}^{s}$ from \eqref{s17} with $\mathbf{W}^s_{n,l}={(\mathbf{Q}^s_{n,l})}^{-1}$.	
	\Repeat
	\State Calculate the matrix $\mathbf{B}^s_{n,l}$ in \eqref{s16} by the obtained $\tilde{\mathbf{x}}_{n,l}^{s}$.
	\State Update the weighting matrix $\mathbf{W}^s_{n,l}$ by the obtained $\mathbf{B}^s_{n,l}$.
	\State Find $\tilde{\mathbf{x}}_{n,l}^{s}$ from (\ref{s17}) with new $\mathbf{W}^s_{n,l}$.
	\Until{convergence}	
	\EndFor
	\State{\bf end}
	\EndFor
	\State{\bf end}
\end{algorithmic}
\end{algorithm}

\vspace{-0.3cm}
\begin{remark}
Ignoring the second- and higher-order noise terms, we yield $\mathbb{E}\{\mathbf{x}\}\approx \mathbf{x}^{\circ}$ and $\mathbb{E}\{\tilde{\mathbf{x}}_{n,l}^{s}\}\approx \tilde{\mathbf{x}}_{n,l}^{s\circ}$.
Thus, the presented
estimator is asymptotically unbiased.
The covariance matrices are given by
$\mbox{\rm cov}(\mathbf{x})\!\approx\!\left((\mathbf{B}^{-1}\mathbf{G})^T\mathbf{Q}^{-1}\mathbf{B}^{-1}\mathbf{G}\right)^{-1}$ 
and
$\mbox{\rm cov}(\tilde{\mathbf{x}}_{n,l}^{s})\!\approx\! \left(({\mathbf{B}^s_{n,l}}^{-1}\mathbf{G}^{s}_{n,l}\mathbf{T})^T{\mathbf{Q}^s_{n,l}}^{-1}{\mathbf{B}^s_{n,l}}^{-1}\mathbf{G}^{s}_{n,l}\mathbf{T}\right)^{-1}$, which approach to their corresponding CRLB under small Gaussian noise levels. Refer to Appendix \ref{D}.
\vspace{-0.3cm}
\end{remark}

\begin{remark}
The weighting matrix in WLS provides the relative importance of the components of an error vector to be minimized \cite{KAY}. 
In the proposed method, the derived weighting matrices ignore the second- and higher-order error terms, which are non-negligible when the noise is large. 
To increase the robustness of the algorithm, the weighting matrices should include the second- and higher-order error components.
An additional refinement mechanism is proposed in the following section to learn higher-order noise terms in a large noise environment by embedding NNs.
\vspace{-0.3cm}
\end{remark}

\section{NN-Assisted WLS Localization}\label{NNWLS}
The model-based WLS estimator proposed in Section \ref{wls1} is proven asymptotically unbiased and effective in achieving the CRLB under small noise conditions.
The general assumption is that the measurement noise follows a Gaussian distribution.
However, in reality, the measurement errors are not completely random. Moreover, an underlying relationship exists between them.
Thus, by utilizing the powerful learning ability of NNs, this underlying relationship can be learned to further improve the localization performance of the proposed WLS estimator, especially at high noise levels.

In this section, we design a NN-assisted WLS (coined as NN-WLS) localization method that embeds NNs into the proposed WLS estimators in \eqref{21} and \eqref{s17}, thereby improving the localization performance. 
Different from treating the NN as a black box (Black Box NN) that directly learns location and velocity,
the NNs in our approach are used to learn the residual vectors $\mathbf{e}$ in \eqref{20} and $\mathbf{e}_{n,l}^s$ in \eqref{s14}, respectively.
Then, the estimated $\mathbf{\hat{e}}$ and $\mathbf{\hat{e}}_{n,l}^s$ are used to construct the weighting matrices $\mathbf{W}$ and $\mathbf{W}_{n,l}^s$ in \eqref{21} and \eqref{s17} and then estimate $\mathbf{x}^{\circ}$ and $\tilde{\mathbf{x}}_{n,l}^{s\circ}$, respectively (Fig. \ref{NN}(c)).
The proposed NN-WLS method can derive more accurate results than the model-based WLS estimator and the Black Box NN method by learning the residual vectors.
We also apply ensemble learning to improve the performance of the proposed NN-WLS method further.

\vspace{-0.25cm}
\subsection{NN-WLS}\label{wlsnet}
As shown in Fig. \ref{NN}(c), the NN-WLS method is a revised version of the WLS estimator derived by introducing learnable vectors $\mathbf{e}$ and $\mathbf{e}_{n,l}^s$.
We provide a general introduction here by taking $\mathbf{e}$ as an example.
According to \cite{KAY}, the weighting matrix is given by $\mathbf{W}=(\mathbb{E}\{\mathbf{e}\mathbf{e}^{T}\})^{-1}$.
In the WLS estimator proposed in Section \ref{wls1}, the vector $\mathbf{e}$ is approximated by the linear term. 
Thus, the approximation error increases with the noise level, thereby compromises the algorithm performance.
Therefore, we propose the NN-WLS method, in which we learn the vector $\mathbf{e}$ by a NN.
The input of the NN is a measurement vector $\mathbf{m}\in \mathbb{R} ^{2(N_a-1)+2N_a}$, which is generated by measurements in set $\mathbb{M}_a$ and given by
\vspace{-0.35cm}
\begin{equation}\label{m}
\mathbf{m}=[r_{21}, \dot{r}_{21}, \ldots, r_{{N_{a}}1}, \dot{r}_{{N_{a}}1},\phi_1, \theta_1, \ldots,\phi_{N_{a}}, \theta_{N_{a}}]^T.
\vspace{-0.35cm}
\end{equation}
Here, the measurement noise is not necessarily Gaussian distributed.
The output of the NN is the estimated residual vector $\hat{\mathbf{e}}$.
Then, the estimated $\hat{\mathbf{e}}$ is used to construct $\mathbf{W}$ by
\vspace{-0.35cm}
\begin{equation}
 \mathbf{W}=(\hat{\mathbf{e}}\hat{\mathbf{e}}^{T}+\epsilon \mathbf{I})^{-1},
 \vspace{-0.35cm}
\end{equation}
where $\epsilon $ is a value to ensure that the inverse of $(\hat{\mathbf{e}}\hat{\mathbf{e}}^{T}+\epsilon \mathbf{I})$ exists.
Finally, we obtain the estimate $\mathbf{x}$ by using the model in \eqref{21}. 
In practice, the training dataset is constructed during an offline phase, in which a site survey is conducted to collect the vectors of the received signals of all RRHs
from different UEs at numerous reference points of known locations, as given in \eqref{receive}. 
Then, the channel parameters are extracted from the received signals with signal processing methods.
The extracted channel parameters construct measurement vector $\mathbf{m}$. Hence, we obtain label $\mathbf{e}$ corresponding to the known location with \eqref{20}. 
Finally, training is performed on the basis of the $T_{train}$ samples, with the structure of each sample as $(\mathbf{m}, \mathbf{e})$.
For simulations,
the location and velocity of the UE are randomly generated for each sample,
then the measurement vector $\mathbf{m}$ is obtained accordingly by \eqref{m},
and $\mathbf{e}$ is generated by \eqref{20}.
We consider the fully connected (FC) NN, 
and the input and output layers both have $4N_{a}-2$ neurons.
The input $(4N_{a}-2)$-dimensional real-valued vector is initially normalized with the value of the element in $[0,1]$.
As for the rectified linear unit (ReLU), $\mbox{ReLU}(x)=\max(x,0)$ is used as the activation function for middle layers.
The sigmoid function $\sigma(x)=1/(1+e^{-x})$ is used as the activation function in the final layer because the output is the normalized vector that has elements scaled within the $[0,1]$ range.
We generate the final estimation $\hat{\mathbf{e}}$ by rescaling. 
The set of parameters is updated by the ADAM algorithm.
The loss function refers to the mean square error (MSE), which is given by
\vspace{-0.25cm}
\begin{equation}\label{loss}
L(\Theta)=\dfrac{1}{T_{train}}\sum_{t=1}^{T_{train}} \Arrowvert \hat{\mathbf{e}}_t - \mathbf{e}_t \Arrowvert ^{2}.
\vspace{-0.25cm}
\end{equation}

Similarly, for the learnable vector $\mathbf{e}_{n,l}^s$,
the input of the NN is a measurement vector $\mathbf{m}_{n,l}^s$, which is generated by measurements in set $\mathbb{M}_{n,l}^s$ and given by
\vspace{-0.25cm}
\begin{equation}\label{m1}
\mathbf{m}_{n,l}^s=[r_{n1,l}^{s}, \dot{r}_{n1,l}^{s}, \phi_{n,l}^{s}, \theta_{n,l}^{s}]^T.
\vspace{-0.25cm}
\end{equation}
The output of the NN is the estimated residual vector $\hat{\mathbf{e}}_{n,l}^{s}$.
Then, the estimated $\hat{\mathbf{e}}_{n,l}^{s}$ is used to construct $\mathbf{W}_{n,l}^{s}$ by
\vspace{-0.25cm}
\begin{equation}
\mathbf{W}_{n,l}^{s}=(\hat{\mathbf{e}}_{n,l}^{s} \hat{\mathbf{e}}_{n,l}^{s T}+\epsilon_{n,l}^{s}\mathbf{I})^{-1},
\vspace{-0.25cm}
\end{equation}
where $\epsilon_{n,l}^{s}$ is a value to ensure the existence of the inverse of $(\hat{\mathbf{e}}_{n,l}^{s} \hat{\mathbf{e}}_{n,l}^{s T}+\epsilon_{n,l}^{s}\mathbf{I})$.
The previously predicted UE location and velocity in vector $\mathbf{x}$ are also used to construct the estimation model \eqref{s17}, by which we 
obtain the estimate $\tilde{\mathbf{x}}_{n,l}^{s}$. 
This part must be executed $\sum_{n=1}^{N}M_n - N_a$ times in parallel to localize all possible scatterers.
The similar FC-NN architecture 
and training process are considered,
except that the input and the output layers have four neurons.

\begin{remark}
	The proposed NN-WLS combines the NNs with the geometric model, thereby consolidating the powerful computing ability of NNs and the robustness of models. The particular advantages are presented as follows. First, the NNs can provide a more accurate estimation of $\mathbf{e}$ and $\mathbf{e}_{n,l}^s$ than the first-order approximation in the previously proposed WLS algorithms. Thus, in some practical scenarios, the NN-WLS can achieve good performance and can be executed even without knowing the covariance matrix $\mathbf{Q}$ and $\mathbf{Q}_{n,l}^s$, whereas the $\mathbf{Q}$ and $\mathbf{Q}_{n,l}^s$ in the WLS algorithms are assumed to be known to initialize the weighting matrix $\mathbf{W}$ and $\mathbf{W}^s_{n,l}$, respectively.
	Moreover, the WLS algorithm is iterative, which implies slow reconstruction, whereas the NN-WLS does not need any iterations, thereby reducing the required time resources.
\end{remark}

\subsection{Ensemble Learning-based NN-WLS}\label{ewlsnet}

\begin{figure}
	\vspace{-0.5cm}
	\centering
	\includegraphics[scale=0.8,angle=0]{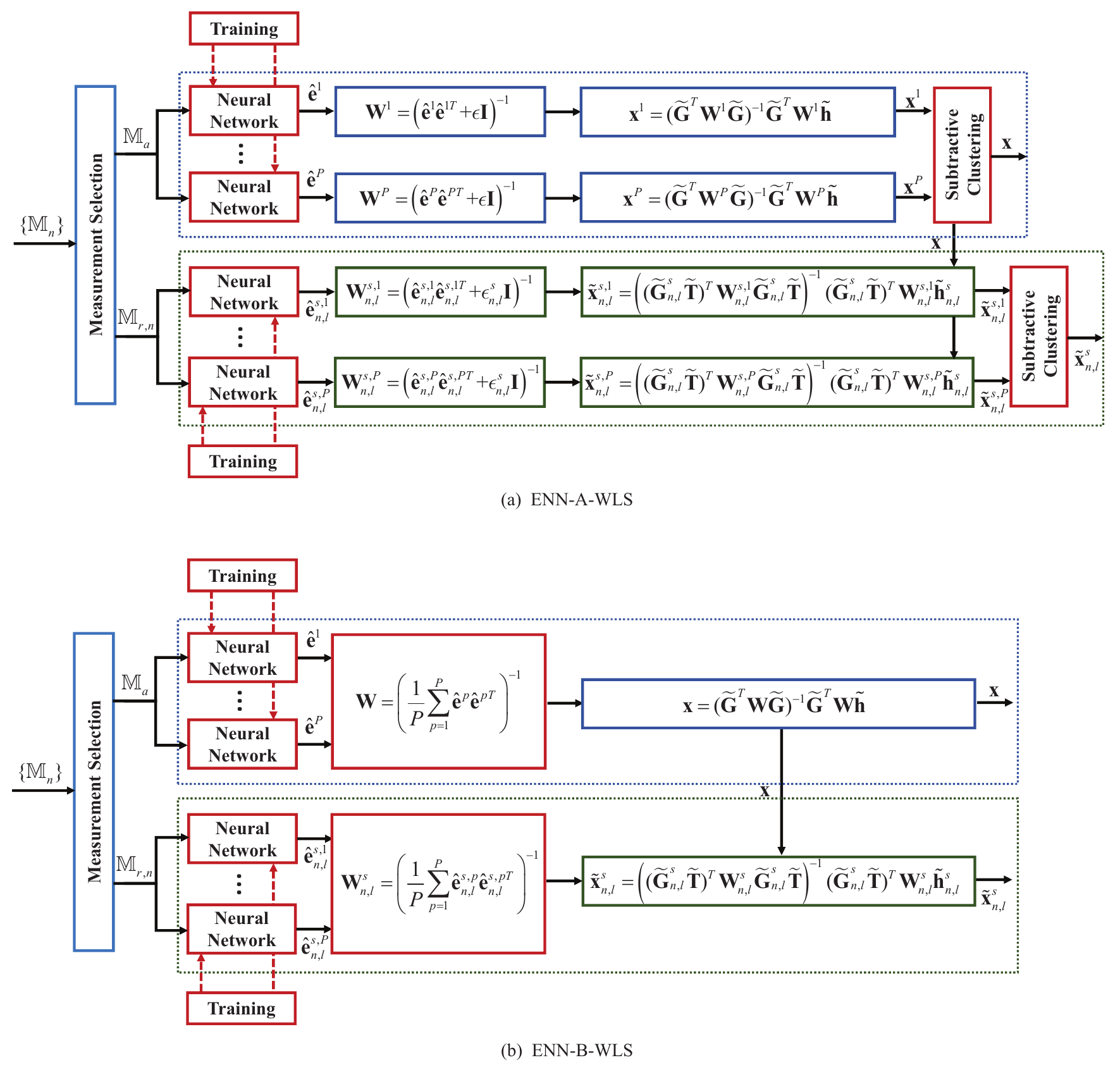}
	\captionsetup{font=footnotesize}
	\caption{Block diagram of two ensemble learning-based NN-WLS localization methods.}
	\label{eWLSNET}
\end{figure}

Training the NN with the loss function defined in \eqref{loss} cannot guarantee that the NN-WLS outputs the globally optimal estimator, even for sufficient data.
According to \cite{ensemble,ensemble2}, the ensemble learning methods often provide additional performance enhancement. 
Ensemble methods correspond to learning algorithms that construct a set of learners and generate a new prediction by taking a vote of the predictions, which may be weighted.
In the backpropagation algorithm for training the NNs, the initial weights of the networks are set randomly.
If the algorithm is applied to the same training dataset but with different initial weights, then the resulting predictions may vary.
NNs that are independently trained with the same training dataset have high probabilities of not making the same prediction error. 
Therefore, we can improve the performance of the NN-assisted WLS algorithm further by introducing an ensemble of $P$-independently trained NNs.

In this study, we propose two ensemble learning-based NN-WLS localization methods, namely, ENN-A-WLS and ENN-B-WLS, as illustrated in Fig. \ref{eWLSNET}(a) and  Fig. \ref{eWLSNET}(b), respectively. 
The following instructions use the localization of UE $\mathbf{x}^\circ$ as an example.
The similarity between the ENN-A-WLS and ENN-B-WLS lies in that 
both of them consist of $P$ independently trained NNs by the same training dataset.
The input of each NN is a measurement vector $\mathbf{m}$ given in \eqref{m} generated by measurements in set $\mathbb{M}_a$,
and the output of each NN is the estimated $\mathbf{\hat e}^p$, for $p = 1,\ldots,P$.
The difference is described as follows:  
As depicted in Fig. \ref{eWLSNET}(a),
the ENN-A-WLS repeats the NN-WLS $P$ times, because $P$ NNs are trained independently and in parallel, 
such that output of each NN-WLS is an independent prediction $\mathbf{x}^p$, for $p = 1,\ldots,P$.
Accurate predictions of UE location are clustered together, and the wrong predictions are located far apart; 
such approach is also applied in UE velocity.
We implement the core part of the ENN-A-WLS, which determines the voting mechanism, by the subtractive clustering.
Unlike the simple averaging method, 
the performance of which seriously deteriorates by the effect of extremely abnormal predictions.
The subtractive clustering method is based on a density measure.
The density measure for the $p$-th location prediction is defined as
\vspace{-0.25cm}
\begin{equation}\label{den}
D_p=\sum\limits_{j=1}^{P}\exp\left(- \lVert \mathbf{u}^p - \mathbf{u}^j \rVert^2 /(r_a/2)^2\right),
\vspace{-0.25cm}
\end{equation}
where $r_a$ is a positive value to denote the radius. The data points outside this radius only contribute slightly to the density measure.
Therefore, by setting a proper $r_a$, the subtractive clustering method can find the point where the predicted values are most clustered.
The point with the highest density measure is selected as the final estimate of UE location.
UE velocity is obtained in the same way.
As shown in Fig. \ref{eWLSNET}(b),
the ENN-B-WLS combines the output $\mathbf{\hat e}^{p}$ of each NN, for $p = 1,\ldots,P$, to construct the weighting matrix as
\vspace{-0.25cm}
\begin{equation}\label{bennwls}
{\mathbf{W}} = {\left( \frac{1}{P}\sum\limits_{p = 1}^P {{{{\hat{\mathbf{e}}}}^{p}}{\hat{\mathbf{e}}^{pT}}}  \right)^{ - 1}},
\vspace{-0.25cm}
\end{equation}
which uses the average of finite $P$ samples to approximate statistical  $\mathbf{W}=(\mathbb{E}\{\mathbf{e}\mathbf{e}^{T}\})^{-1}$.
Then, we obtain the estimate $\mathbf{x}$ by using the model in \eqref{21} with the constructed ${\mathbf{W}}$ in \eqref{bennwls}. 
Scatterers are localized in a similar way and further details are omitted.

\begin{table}[t]
	\vspace{0.8cm}	
	\renewcommand{\arraystretch}{1.5}
	\centering
	\fontsize{8}{8}\selectfont
	\captionsetup{font=small}
	\caption{Locations of the RRHs in meters.}\label{Tbs}
	\begin{threeparttable}	
		\begin{tabular}{cccccccccc}
			\toprule
			& 1 & 2 & 3 & 4 & 5 & 6 & 7 & 8 & 9\\
			\hline
			x & 235.5042 & 287.5042 & 235.5042 & 287.5042 &  235.5042 & 287.5042 & 235.5042 & 287.5042 & 235.5042 \\
			y & 389.5038 & 389.5038 & 489.5038 & 489.5038 &  589.5038 & 589.5038 & 851.5038 & 851.5038 & 651.5038\\
			z & 26 & 32 & 10 & 40 & 14 & 50 & 26 & 26 & 26 \\      		
			\bottomrule
			\toprule
			& 10 & 11 & 12 & 13 & 14 & 15 & 16 & 17 & 18 \\
			\hline
			x & 287.5042 & 235.5042 & 287.5042 & 235.5042 & 287.5042 & 235.5042 & 287.5042 & 235.5042 & 287.5042 \\		
			y & 651.5038 & 751.5038 & 751.5038 & 851.5038 & 851.5038 & 951.5038 & 951.5038 & 1051.5038 & 1051.5038 \\
			z & 26 & 26 & 26 & 26 & 26 & 26 & 26 & 26 & 26 \\
			\bottomrule
		\end{tabular}
	\end{threeparttable}
	\vspace{-1cm}
\end{table}
\section{Numerical Results}

\subsection{Model-based WLS Localization}\label{wls}
In this subsection, we analyze the performance of the proposed WLS estimator.
We consider a scenario with $N=18$ RRHs, and their locations are given in Table \ref{Tbs}. 
\footnote{	
Our proposed method (i.e., TDOA/FDOA/AOA) can work with either a randomized selection of z-axis coordinates for different RRHs or the same value of z-axis coordinates for all RRHs. To compare the performance of different methods, we choose a randomized selection of the z-axis coordinates for the first six RRHs.}
The UE is located at $\mathbf{u}^\circ=[250,450,0]^T$ m with the velocity $\dot{\mathbf{u}}^\circ=[-10,2,5]^T$ m/s.
The CU selects $N_{a}$ LOS paths from RRHs to locate the UE.
Although the presented algorithm is derived for Gaussian noise model with general covariance matrix, 
we consider the following form of the covariance matrix of the noise terms $\Delta \mathbf{m}$ for simplicity,
\vspace{-0.25cm}
\begin{equation}\label{mm}
\mathbf{Q}=\mbox{blkdiag}(\overbrace{\mathbf{Q}_d,\ldots,\mathbf{Q}_d}^{(N_{a}-1)},
\overbrace{\mathbf{Q}_a,\ldots,\mathbf{Q}_a}^{N_{a}}),
\vspace{-0.25cm}
\end{equation}
where  $\mathbf{Q}_d=\mbox{diag}(\delta_d^2,(0.1\delta_d)^2)$,
$\mathbf{Q}_a=\mbox{diag}(\delta_a^2,\delta_a^2)$, and $\delta_d$, $0.1\delta_d$, and $\delta_a$ are the standard deviations of TDOA, FDOA, and AOA measurements.
The order of the elements in \eqref{mm} is the same as that in $\Delta \mathbf{m}$, in which the first $(N_{a}-1)$ pairs are TDOA and FDOA pairs (the covariance matrix for each pair is $\mathbf{Q}_d$), and the last $N_{a}$ pairs are AOA pairs (the covariance matrix for each pair is $\mathbf{Q}_a$).
Similarly, we consider the covariance matrix of the $\Delta\mathbf{m}_{n,l}^{s}$ for the $(n,l)$-th scatterer in the form of 
$\mathbf{Q}^s_{n,l}=\mbox{diag}(\delta_d^2,(0.1\delta_d)^2,\delta_a^2,\delta_a^2)$.
The localization accuracy is
assessed via the root mean square error (RMSE), e.g., $\mbox{RMSE}(\mathbf{u})=\sqrt{\sum_{t=1}^{T_{MC}}||\mathbf{u}_t-\mathbf{u}^\circ||^2/T_{MC}}$,
where $\mathbf{u}_t$ is
the estimate of $\mathbf{u}^\circ$ at the $t$-th Monte Carlo simulation.

\begin{figure}
	\vspace{-0.25cm}
	\begin{minipage}{0.48\textwidth}
		\centering
		\captionsetup{font=footnotesize}
		\includegraphics[scale=0.6,angle=0]{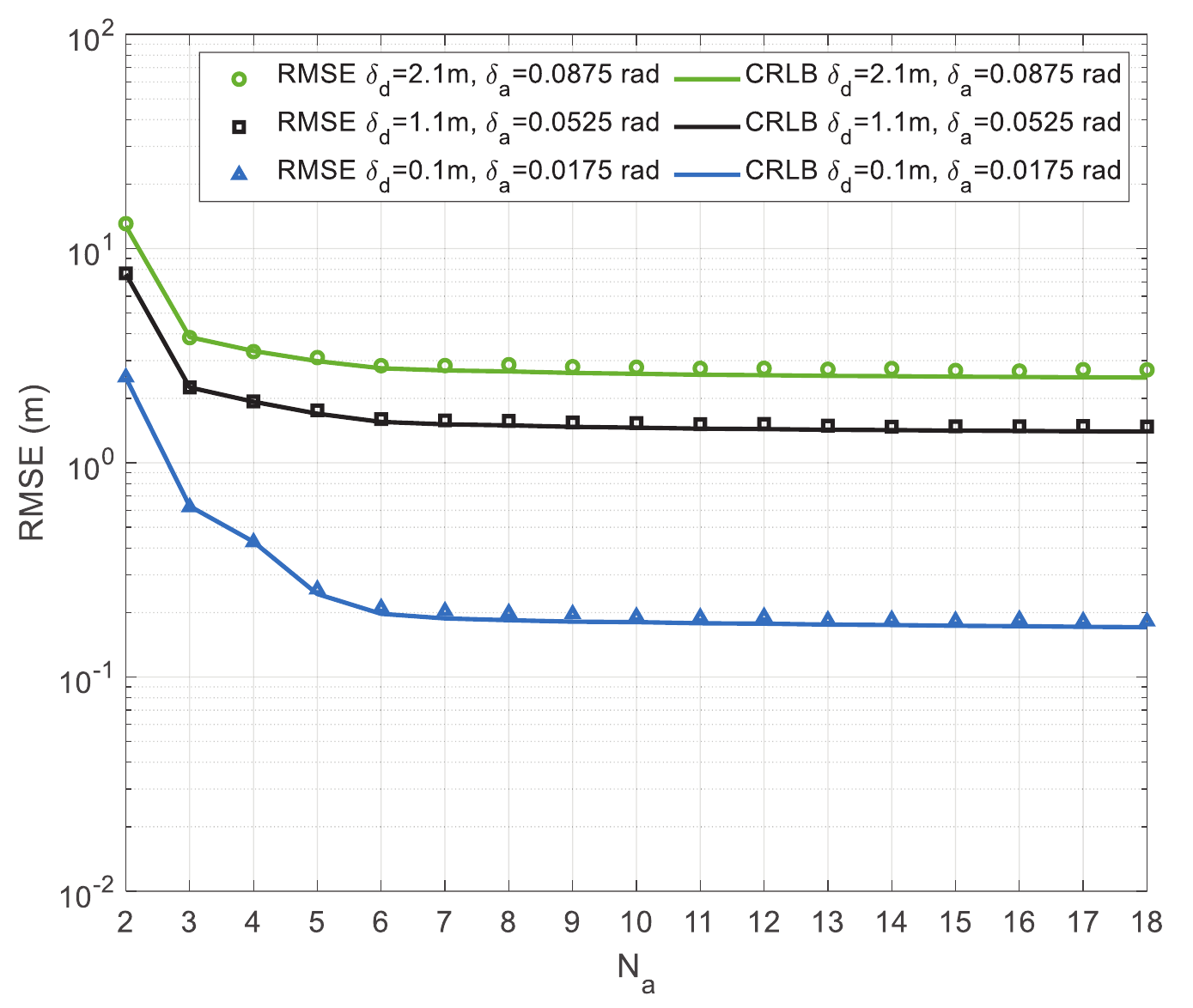}
		\caption{RMSE performance of the proposed algorithm in location estimation with different numbers of selected LOS paths.}
		\label{RRH_u}
	\end{minipage}
	\hspace{0.5cm}
	\begin{minipage}{0.48\textwidth}
		\centering
		\captionsetup{font=footnotesize}
		\includegraphics[scale=0.6,angle=0]{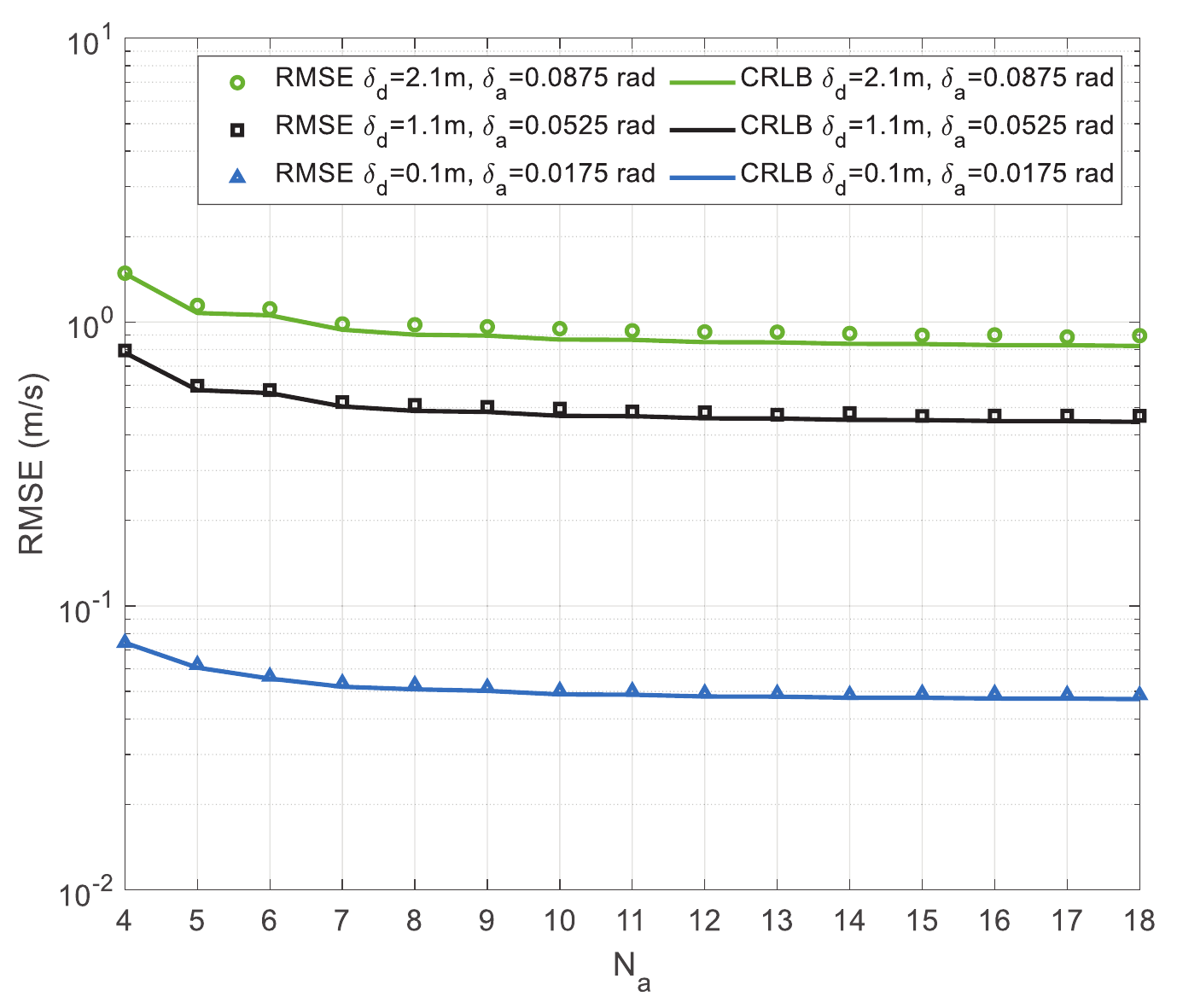}
		\caption{RMSE performance of the proposed algorithm in velocity estimation with different numbers of selected LOS paths.}
		\label{RRH_v}
	\end{minipage}
\end{figure}

In the first simulation scenario, we evaluate the performance of the proposed UE localization algorithm with different numbers of selected LOS paths.
Fig. \ref{RRH_u} and Fig. \ref{RRH_v} depict the RMSEs versus  $N_a$.
Here, the numerical results are obtained from $T_{MC}=5000$ independent Monte Carlo simulations.
Note that having a larger number of LOS paths is beneficial to achieve localization accuracy.
For location estimation (Fig. \ref{RRH_u}), 
the proposed WLS algorithm requires $N_a \geqslant 2$ LOS paths.
The localization accuracy is significantly enhanced as $N_a$ increases to $3$ and is saturated when $N_a \geqslant 6$.
For velocity estimation (Fig. \ref{RRH_v}), 
the proposed WLS estimator requires $N_a \geqslant 4$ LOS paths.
The performance improves gradually with $N_a$ and reaches saturation for $N_a \geqslant 6$.
In all cases, the CRLBs can be attained,
and the bounds are tighter for smaller $N_a$, $\delta_d$, and $\delta_a$.
These results demonstrate that as long as $4$-$6$ LOS paths are available, the proposed algorithm can realize UE localization  with acceptable performance.

\begin{figure}[H]
	\vspace{-0.25cm}
	\centering
	\captionsetup{font=footnotesize}
	\includegraphics[scale=0.65,angle=0]{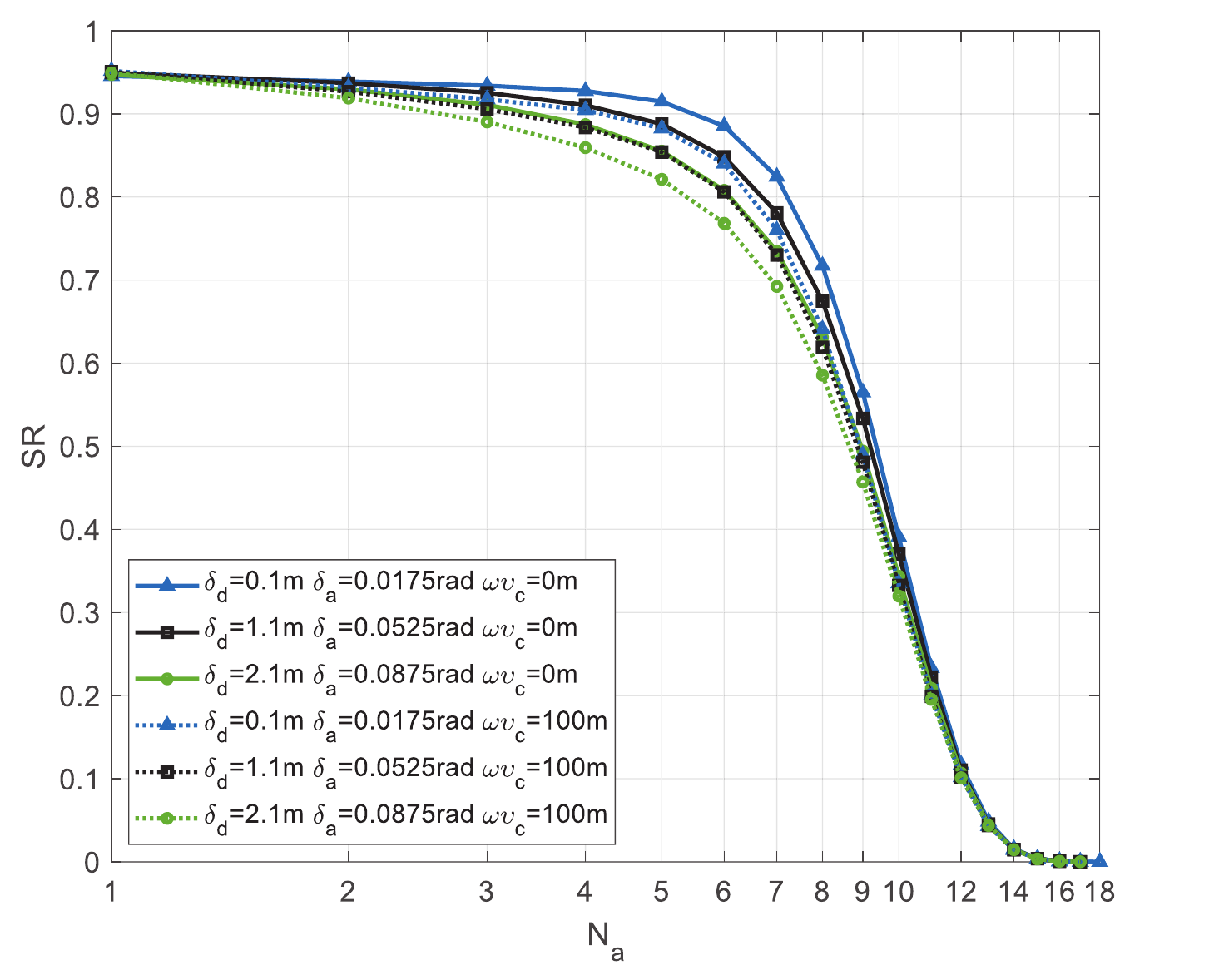}
	\caption{SR performance of the proposed measurement selection method.}
	\label{SR}
\end{figure}

In the second simulation scenario, 
we want to reveal that four to six LOS paths can be obtained in mmWave CRAN communication systems by densely deploying RRHs and designing appropriate measurement selection methods.
We analyze the performance of the proposed measurement selection method by using the following simulation settings.
The detection probability $P_{d}$ for each RRH is set to $0.5$.
Each scatterer is distributed uniformly in a 3-D space
$\{[x, y, z]^T:240 \leqslant x \leqslant 280, 450 \leqslant y \leqslant 850, 0 \leqslant z \leqslant 20 \}$ in meters,
whilst the magnitude of velocity follows $\mathcal U [0,10]$ m/s.
Successful selection is the phenomenon in which all of
the selected $N_a$ paths are LOS paths. Thus, the success rate (SR) $T_{SR}/T_{MC}$ signifies that $T_{SR}$ times successful selection out of $T_{MC}$ Monte Carlo simulations, and we set $T_{MC} = 100000$.
TOA and AOA measurements used in this study follow Gaussian distributions with mean given by \eqref{1} and \eqref{5}, respectively, and standard deviations given by $\delta_d$ and $\delta_a$, respectively.
Fig. \ref{SR} shows the SR performance versus $N_a$
 by setting (1) $\delta_d = 0.1$ m and $\delta_a = 0.0175$ rad, (2) $\delta_d = 1.1$ m and $\delta_a = 0.0525$ rad, (3) $\delta_d = 2.1$ m and $\delta_a = 0.0875$ rad. 
 ($0.0175$ rad $=1^\circ $,
 $0.0525$ rad $=2^\circ $, 
 $0.0875$ rad $=3^\circ $, respectively). 
We also considered clock bias setting $\omega\upsilon_c=0$ m and $\omega\upsilon_c=100$ m for each noise level configuration.
In all cases, the SR achieves $85\%$ when $N_a = 4$, and the SR is larger than $80\%$ for $N_a \leqslant 6$ in most cases.
The SR can be further improved by increasing the detection probability of RRH and by considering the energy gap between the LOS and NLOS paths \footnote{ For example, let $P_{max}$ denote the energy of the strongest path, set the threshold as $P_{thre}=P_{max}/2$, and the paths with energy less than $P_{thre}$ are filtered out.
Note that $P_{max}/2$ is an empirical setting based on a general ray-tracing dataset for mmWave massive MIMO \cite{deepmimo}. }.
However, this topic is not the focus of this study, hence, we will not go into further details.

\begin{figure}
	\vspace{-0.25cm}
	\begin{minipage}[t]{0.48\textwidth}
		\centering
		\captionsetup{font=footnotesize}
		\includegraphics[scale=0.6,angle=0]{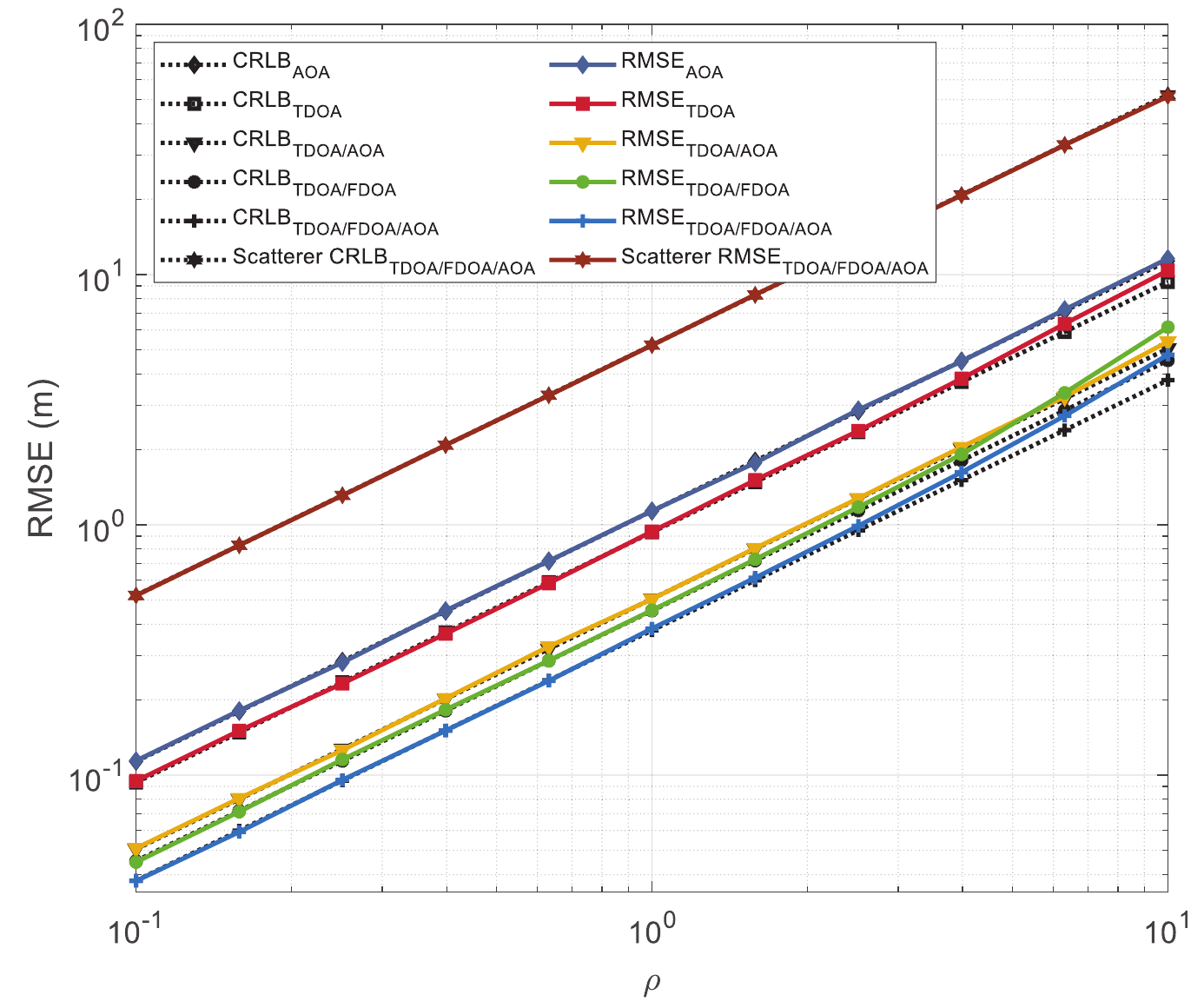}
		\caption{Comparison of the RMSE of the proposed algorithm in location estimation with that of the AOA-only, TDOA-only, TDOA/AOA, TDOA/FDOA algorithms, and the corresponding CRLBs.}
		\label{CRLB_u}
	\end{minipage}
	\hspace{0.5cm}
	\begin{minipage}[t]{0.48\textwidth}
		\centering
		\captionsetup{font=footnotesize}
		\includegraphics[scale=0.6,angle=0]{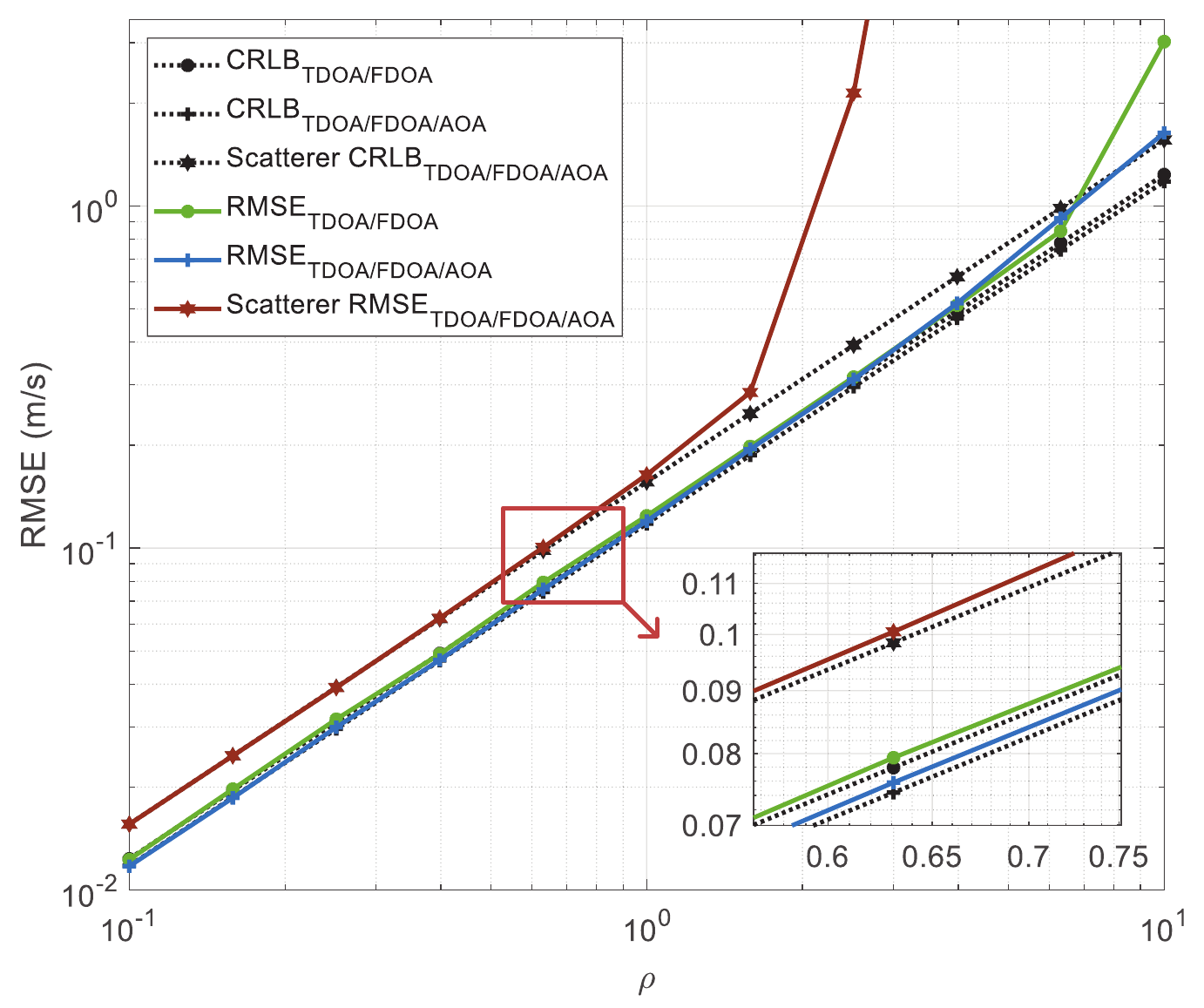}
		\caption{Comparison of the RMSE of the proposed algorithm in velocity estimation with that of the AOA-only, TDOA-only, TDOA/AOA, TDOA/FDOA algorithms, and the corresponding CRLBs.}
		\label{CRLB_v}
	\end{minipage}
\end{figure}

In the third simulation scenario, we evaluate the performance of the proposed TDOA/FDOA/AOA based WLS estimator by comparing it with AOA-only, TDOA-only\cite{TD3}, TDOA/AOA\cite{l2}, TDOA/FDOA\cite{l3} WLS estimators, and the corresponding CRLBs.
We set $N_{a} = 6$, $T_{MC}=5000$, $\delta_d = 0.22 \rho$, and $\delta_a = 0.0175\rho$, where $\rho$ is a noise scaling factor (See Table \ref{bias} for specific values).
The RMSEs and CRLBs of different estimators are shown in Fig. \ref{CRLB_u} and Fig. \ref{CRLB_v} as
functions of the noise scaling factor.
The results in Fig. \ref{CRLB_u} show that the proposed TDOA/FDOA/AOA based WLS estimator has the best performance, followed by TDOA/FDOA, TDOA/AOA, TDOA-only, and AOA-only.
For velocity estimation in Fig. \ref{CRLB_v}, 
we only compare the proposed estimator with the TDOA/FDOA WLS estimator because velocity cannot be obtained without FDOA measurements.
Fig. \ref{CRLB_v} shows that the performance of the proposed estimator is slightly remarkable.
Furthermore, the proposed TDOA/FDOA/AOA localization can achieve the CRLB for small noise level.
Increasing the noise level results in a slow deviation from the CRLB for both location and velocity estimations because the nonlinear terms in $\mathbf{e}$ in the derivation of the proposed algorithm have been ignored.
TDOA/FDOA algorithm uses two-stage WLS estimators and has larger deviation from CRLB than the proposed estimator as the noise level increases.
The proposed scatterer localization performance is also depicted in 
Fig. \ref{CRLB_u} and Fig. \ref{CRLB_v}.
The unknown scatterer is located at $[240,600,-19]^T$ in meters.
The velocity direction of the scatterer is the same as $\dot{\mathbf{u}}^\circ$ with a magnitude of $5$ (m/s).
The results in Fig. \ref{CRLB_u} demonstrate that, for the scatterer location, the RMSE can achieve the CRLB. 
However, the CRLB of the scatterer is higher than that of the UE because the number of measurements used in scatterer localization is less than that in the UE localization.
For the scatterer velocity observed in Fig. \ref{CRLB_v},
the RMSE can achieve the CRLB when $\rho \leqslant 1$.
Since velocity is mainly determined by the FDoA measurements, and only one measurement can be used for each scatterer, 
the proposed algorithm can ensure good performance with relatively small noise and is greatly affected by large noise caused by insufficient measurements.

\subsection{NN-assisted WLS Localization}
\begin{table}
	\vspace{1cm}
	\centering
	\begin{minipage}[b]{0.48\textwidth}
		\renewcommand{\arraystretch}{1.5}
		\centering
		\fontsize{8}{8}\selectfont
		\captionsetup{font=small}
		\caption{Measurement noise settings.}\label{bias}
		\begin{threeparttable}	
			\begin{tabular}{cccc}
				\toprule
				$\rho$ & 0.1 & 1 & 10 \\
				\hline
				$\delta_d$ (m) & 0.022 & 0.22 & 2.2 \\
				$\delta_a$ (rad) & 0.00175 & 0.0175 & 0.175 \\		
				\bottomrule			
			\end{tabular}
		\end{threeparttable}
	\end{minipage}
	\hspace{0.15cm}
	\begin{minipage}[b]{0.48\textwidth}
		\renewcommand{\arraystretch}{1.5}
		\centering
		\fontsize{8}{8}\selectfont
		\captionsetup{font=small}
		\caption{MAE performance comparison.}
		\label{tab:performance_comparison}
		\begin{threeparttable}			
			\begin{tabular}{ccc}
				\toprule
				{Method}& Location (m) & Velocity (m/s) \cr		
				\midrule
				Black Box & 0.1782 & 0.2109 \cr
				WLS & 0.0200 & 0.0143  \cr			
				NN-WLS & {\bf 0.0104} & {\bf 0.0054}  \\								
				\bottomrule
			\end{tabular}
		\end{threeparttable}
	\end{minipage}
	\vspace{-1cm}
\end{table}

In this subsection, we explore the performance of the proposed NN-assisted WLS localization methods. 
First,
we utilize a general dataset for mmWave massive MIMO constructed on the basis of the ray-tracing data from Remcom Wireless InSite \cite{deepmimo} to verify the effectiveness of the proposed algorithms because this approach can simulate real-world scenarios accurately. 
Specifically, we evaluate the performance of the Black Box NN (Section \ref{BBNN}), the proposed WLS (Section \ref{wls1}), and the proposed NN-WLS (Section \ref{wlsnet}) through the same revised ray-tracing dataset.\footnote{We utilize the first $6$ RRHs in the ray-tracing dataset, and
	each sample is generated with a different UE location distributed in a 3D space
	$\{[x, y, z]^T:240 \leqslant x \leqslant 280, 410 \leqslant y \leqslant 740, z = 2 \}$ in meters. 
	There are no FDoA measurements given in the ray-tracing dataset. For each UE, we generate its velocity in a random way, and then calculate its corresponding FDoA measurements.}
The training, validation, and testing datasets contain $60000$, $20000$, and $20000$ samples, respectively. 
All testing samples are excluded from the training and validation samples.
	The inner architecture of the networks used for the Black Box NN and proposed NN-WLS is identical and consists of a three-layer FC-NN.
	The first two FC layers use 32 neurons, and the third FC layer uses 22 neurons.
The localization accuracy is assessed via the mean absolute error (MAE), e.g., $\mbox{MAE}(\mathbf{u})={\sum_{t=1}^{T_{test}}||\mathbf{u}_t-\mathbf{u}_t^\circ||/T_{test}}$,
where $\mathbf{u}_t$ is the estimation of $\mathbf{u}_t^\circ$ in the test dataset, and $T_{test}$ is the size of test dataset.
The MAE results of the Black Box NN, WLS, and NN-WLS are given in Table \ref{tab:performance_comparison}.
The result shows that the NN-WLS is more accurate in terms of location and velocity estimation than the WLS algorithm.
The Black Box NN is the simplest to operate but has the worst accuracy.
The results verify that the measurement errors are not completely random, that is, an underlying relationship exists between them, and this relationship can be learned by the NN, which motivates our research.
For an in-depth analysis, we explore the performance of the proposed NN-assisted WLS localization methods with different noise levels and training dataset sizes in the following.

\subsubsection{Localization Accuracy to Noise Level}
\begin{table}[t]
	\vspace{1cm}	
	\renewcommand{\arraystretch}{1.5}
	\centering
	\fontsize{8}{8}\selectfont
	\captionsetup{font=small}
	\begin{threeparttable}
		\caption{MAE (m) performance comparison of UE location estimation.}
		\label{tab:performance_comparison_u}
		\begin{tabular}{cccccccccc}
			\toprule
			\multirow{2}{*}{Method}&
			\multicolumn{3}{c}{ $\delta_d=3 $ m, $\delta_a=0.0525$ rad}&\multicolumn{3}{c}{ $\delta_d=3 $ m, $\delta_a=0.0175$ rad}&\multicolumn{3}{c}{ $\delta_d=0.1 $ m, $\delta_a=0.0525$ rad}\cr
			\cmidrule(lr){2-4} \cmidrule(lr){5-7} \cmidrule(lr){8-10}
			&Ratio=0.1&Ratio=0.01&Ratio=0.001&Ratio=0.1&Ratio=0.01&Ratio=0.001&Ratio=0.1&Ratio=0.01&Ratio=0.001\cr
			\midrule
			WLS & 11.42 & 11.23 & 11.17 & 3.86 & 3.85 & 3.85 & 2.12 & 2.10 & 2.10 \cr
			Black Box & 3.44 & 1.72 & 2.92 & 3.38 & 2.06 & 1.54 & 3.03 & 1.23 & 1.42 \cr
			NN-WLS & {\bf 2.58} & {\bf 0.55} & {\bf 0.16} & {\bf 2.24} & {\bf 0.55} & {\bf 0.09} & {\bf 0.25} & {\bf 0.05} & {\bf 0.03} \cr						
			\bottomrule
		\end{tabular}
	\end{threeparttable}
	\vspace{0.5cm}
\end{table}
\begin{table}[t]
	\renewcommand{\arraystretch}{1.5}	
	\centering
	\fontsize{8}{8}\selectfont
	\captionsetup{font=small}
	\begin{threeparttable}
		\caption{MAE (m/s) performance comparison of UE velocity estimation.}
		\label{tab:performance_comparison_v}
		\begin{tabular}{cccccccccc}
			\toprule
			\multirow{2}{*}{Method}&
			\multicolumn{3}{c}{ $\delta_d=3 $ m, $\delta_a=0.0525$ rad}&\multicolumn{3}{c}{ $\delta_d=3 $ m, $\delta_a=0.0175$ rad}&\multicolumn{3}{c}{ $\delta_d=0.1 $ m, $\delta_a=0.0525$ rad}\cr
			\cmidrule(lr){2-4} \cmidrule(lr){5-7} \cmidrule(lr){8-10}
			&Ratio=0.1&Ratio=0.01&Ratio=0.001&Ratio=0.1&Ratio=0.01&Ratio=0.001&Ratio=0.1&Ratio=0.01&Ratio=0.001\cr
			\midrule
			WLS & 3.12 & 2.98 & 2.83 & 2.88 & 2.80 & 2.79 & 0.18 & 0.18 & 0.18 \cr
			Black Box & 1.29 & 1.39 & 1.20 & 1.27 & 1.12 & 1.11 & 1.18 & 1.00 & 1.09 \cr
			NN-WLS & {\bf 0.68} & {\bf 0.25} & {\bf 0.22} & {\bf 0.59} & {\bf 0.15} & {\bf 0.13} & {\bf 0.12} & {\bf 0.07} & {\bf 0.06} \cr 			
			\bottomrule
		\end{tabular}
	\end{threeparttable}
	\vspace{-1cm}
\end{table}
We increase the noise level of measurements 
to analyze the performance of the NN-assisted WLS algorithm.
By observing the ray-tracing dataset, we find that the measurement errors include a dominant part and a fluctuating part. 
We define the dominant part as the unknown fixed error and the fluctuating part as the Gaussian random error. 
We define three dominant error settings: 
(1) $\delta_d=3 $ m, $\delta_a=0.0525$ rad; (2) $\delta_d=3 $ m, $\delta_a=0.0175$ rad; (3) $\delta_d=0.1 $ m, $\delta_a=0.0525$ rad.
Three radios are available for each setting, 
and the standard deviation of the fluctuating error are $0.1$, $0.01$, and $0.001$ times of that of the dominant error. 
Therefore, nine noise settings have been identified.
Training and testing are conducted under the same noise setting.
The training, validation, and testing sets contain $12000$, $4000$, and $4000$ samples, respectively. 
All testing samples are excluded from the training and validation samples.
The MAE results for different methods are shown in Table \ref{tab:performance_comparison_u} and Table \ref{tab:performance_comparison_v}.
The performance of the proposed NN-WLS outperforms the WLS algorithm and the black box NN in the given simulation scenarios.
Moreover, 
by decreasing the ratio of the error standard deviation of the random part to that of the fixed part,
the MAE of NN-WLS and black box NN decreases.
That is, 
as the proportion of the random part decreases, the ability of the NNs increases.
This is due to the fact that the NNs can learn the dominant error and the correlation between measurement errors, but WLS algorithm cannot.

\subsubsection{Network Performance to Training Dataset Size}

\begin{figure}[t]
	\vspace{-0.25cm}
	\begin{minipage}[t]{0.48\textwidth}
		\centering
		\captionsetup{font=footnotesize}
		\includegraphics[scale=0.6,angle=0]{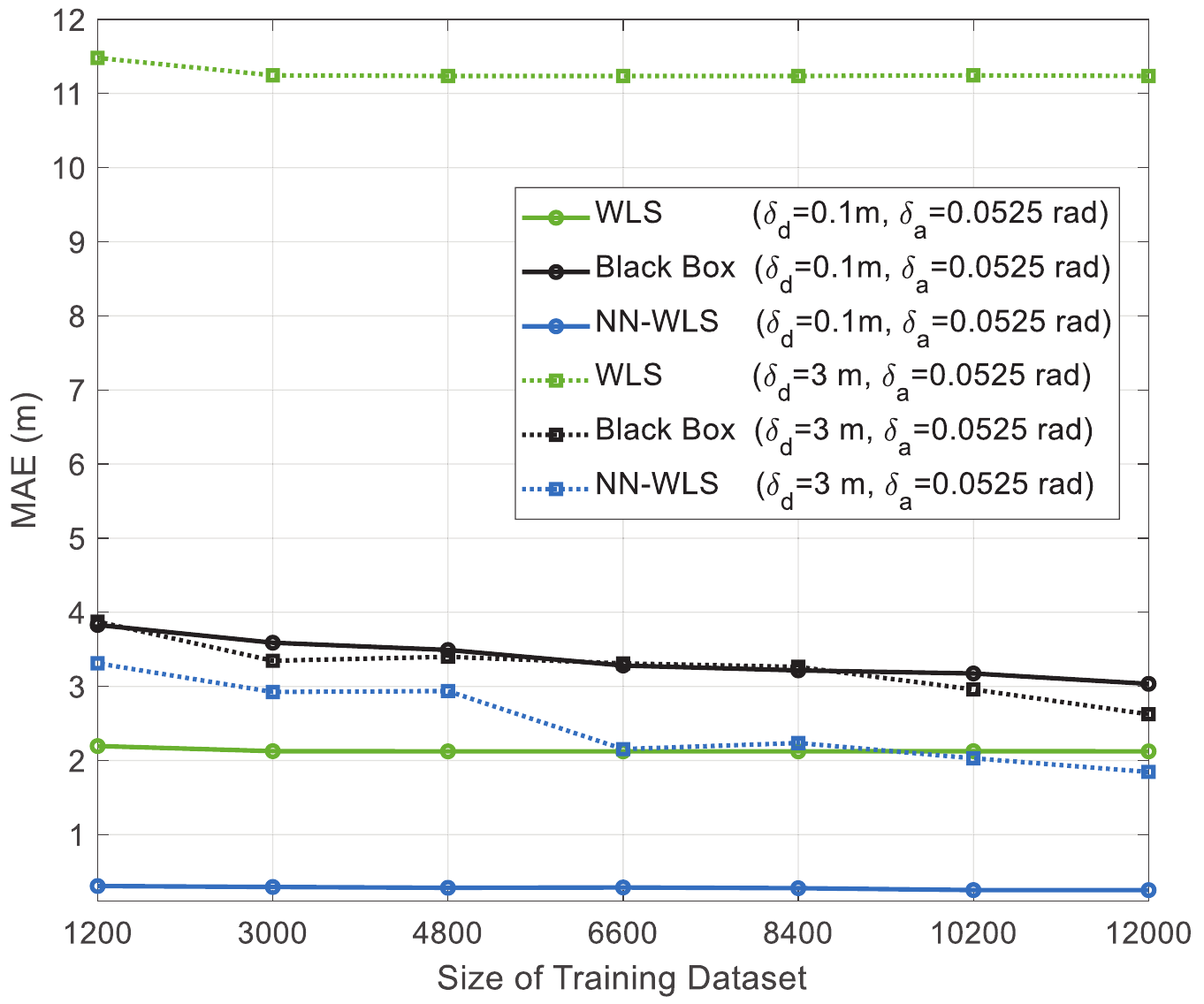}
		\caption{MAE performance comparison of location estimation between the black box and NN-WLS algorithms for varying training dataset sizes.}
		\label{reduce_u}
	\end{minipage}
	\hspace{0.5cm}
	\begin{minipage}[t]{0.48\textwidth}
		\centering
		\captionsetup{font=footnotesize}
		\includegraphics[scale=0.6,angle=0]{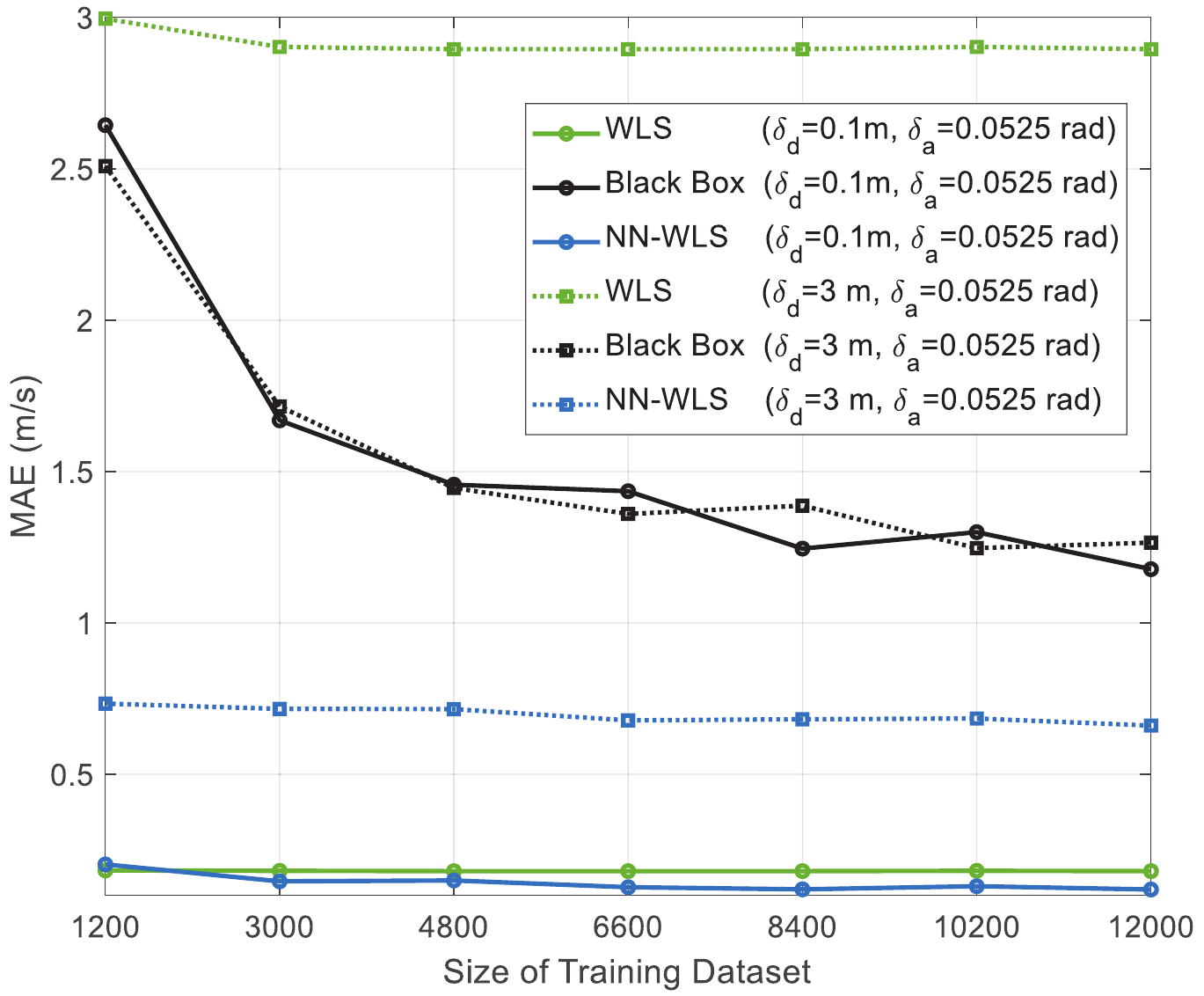}
		\caption{MAE performance comparison of velocity estimation between the black box and NN-WLS algorithms for varying training dataset sizes.}
		\label{reduce_v}
	\end{minipage}
\end{figure}

We reduce the size of training dataset from $12000$ to $1200$, and the performance of the NN-WLS and Black Box is shown in Fig. \ref{reduce_u} and  Fig. \ref{reduce_v}. In all simulations, the ratio is set to $0.1$.
For a relatively large noise level, where $\delta_d=3$ m and $\delta_a=0.0525$ rad, 
the performance of NN-WLS saturates with $6600$ and $1200$ training data for location and velocity estimation, respectively, whereas the black box NN requires more training data to increase accuracy.
In addition, reducing the noise level can bring gains to NN-WLS, but not to the black box NN, since the latter is purely data driven and lacks the assistance of the geometric model.
When $\delta_d=0.1$ m and $\delta_a=0.0525$ rad, NN-WLS adds $1200$ and $3000$ training samples on the basis of WLS, the estimation accuracy of UE location and velocity can be improved by $86\%$ and $19\%$, respectively. 
When $\delta_d=3$ m and $\delta_a=0.0525$ rad, NN-WLS adds $6600$ and $1200$ training samples, the estimation accuracy of UE location and velocity can be improved by $88\%$ and $76\%$, respectively.
The performance of the proposed WLS estimator is enhanced by the NN, especially in a large noise environment. 

\subsubsection{Network Robustness to Measurement Noise}
\begin{figure}[t]
	\centering
	\captionsetup{font=footnotesize}
	\includegraphics[scale=0.9,angle=0]{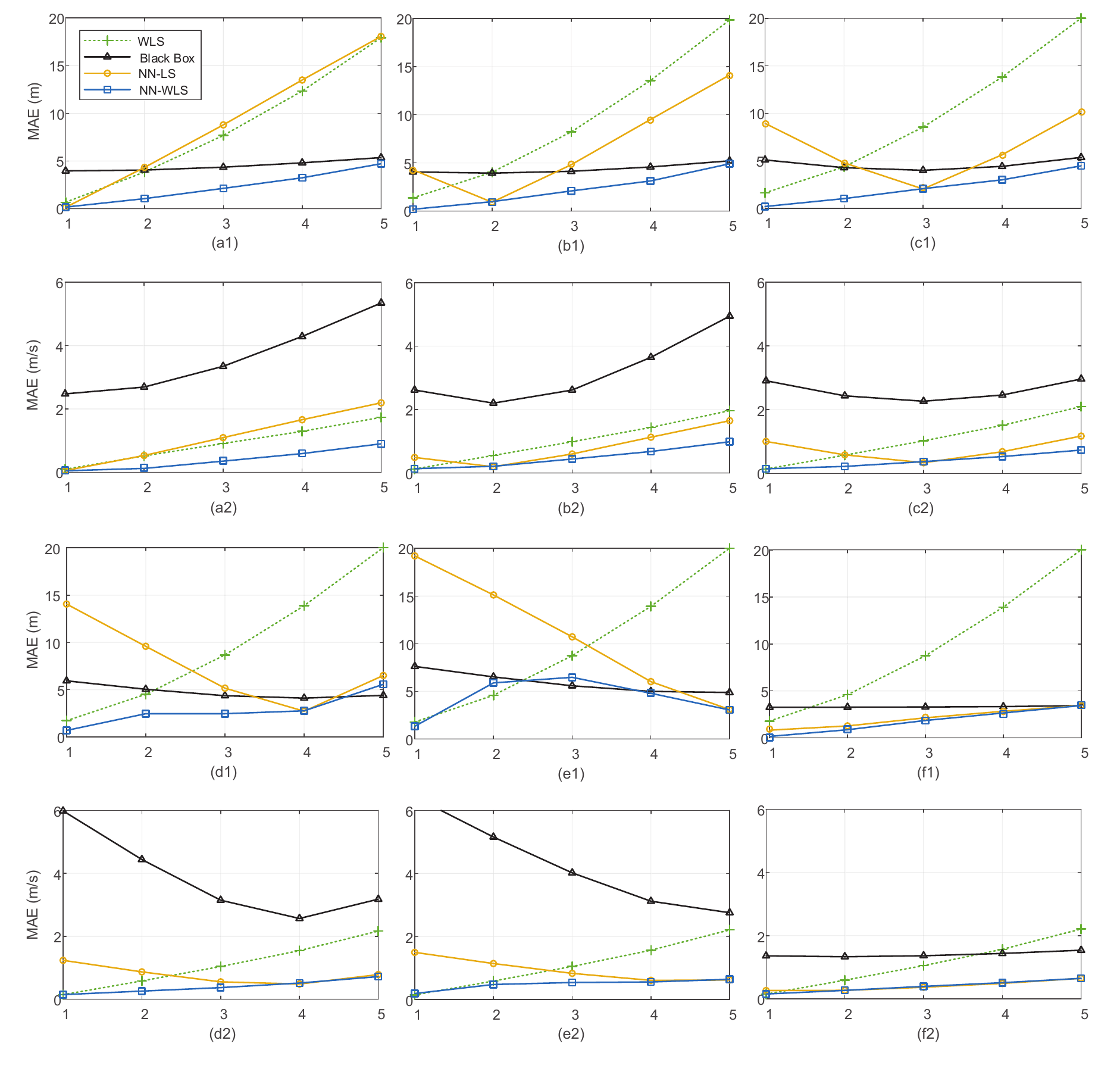}
	\caption{MAE performance comparison among the WLS, black box, NN-WLS, and NN-LS algorithms for various noise conditions.}
	\label{robust}
\end{figure}
We study the robustness of the proposed NN-WLS to the varying measurement noise conditions. 
For comparison, we define the NN-LS algorithm. In particular, after obtaining the estimated residual vector $\hat{\mathbf{e}}$ from the NN (the same way as that implemented in the NN-WLS), we deduct $\hat{\mathbf{e}}$ from \eqref{20}. Then, by directly applying the LS algorithm, we obtain
$
\mathbf{x}=(\tilde{\mathbf{G}}^{T}\tilde{\mathbf{G}})^{-1}\tilde{\mathbf{G}}^{T}(\tilde{\mathbf{h}} - \hat{\mathbf{e}}).
$
The black box NN, NN-WLS, and NN-LS are executed using the same datasets.
Fig. \ref{robust} illustrates the performance
of the black box NN, NN-WLS, and NN-LS trained for a specific noise level and deployed in different noise levels.
We have five different measurement noise settings:
(1) $\delta_d=0.1 $ m and $\delta_a=0.0175$ rad,
(2) $\delta_d=0.6 $ m and $\delta_a=0.035$ rad,
(3) $\delta_d=1.1 $ m and $\delta_a=0.0525$ rad,
(4) $\delta_d=1.6 $ m and $\delta_a=0.07$ rad,
(5) $\delta_d=2.1 $ m and $\delta_a=0.0875$ rad.
In all simulations, the ratio is set to $0.1$.
The black box NN, NN-WLS, and NN-LS in Figs. \ref{robust}(a1) and (a2) are trained in the noise setting (1) and are tested in noise settings (1) to (5),
where Fig. \ref{robust}(a1) shows the MAE performance of location estimation and Fig. \ref{robust}(a2) shows the MAE performance of velocity estimation.
Figs. \ref{robust}(b1) and (b2),  Figs. \ref{robust}(c1) and (c2),
Figs. \ref{robust}(d1) and (d2), and Figs. \ref{robust}(e1) and (e2) are trained in noise settings (2), (3), (4), and (5), respectively.
The size of the training dataset is $1200$ in Figs. \ref{robust}(a1) and (a2) to Figs. \ref{robust}(e1) and (e2). 
Moreover, the training dataset in Figs. \ref{robust}(f1) and (f2) includes all (1)-(5) measurement noise settings, and the size of the training dataset is $8000$.

The results indicate that NN-WLS is robust for small noise settings and 
outperforms the black box NN in most cases in terms of location estimation.
In addition,
NN-WLS is robust for all the noise settings in terms of velocity estimation, whereas the black box NN shows great performance fluctuations.
When the noise setting of test dataset is the same as that of the training dataset,
the performance of NN-LS is comparable to NN-WLS.
However, in terms of both location and velocity estimation, NN-LS
performs poorly when tested by using a different noise setting from the training dataset. 
NN-LS requires the estimated $\hat{\mathbf{e}}$ to be highly accurate, so that the LS algorithm can be used to derive good results. 
By contrast, in NN-WLS, the weighting matrix is
$\mathbf{W}=(\hat{\mathbf{e}}\hat{\mathbf{e}}^{T}+\epsilon\mathbf{I})^{-1}$, which contains the information of the dominant ($\hat{\mathbf{e}}$ is the learned mean of the dominant error) and the random error parts ($\epsilon\mathbf{I}$ is the covariance matrix of the Gaussian random error).\footnote{ In the simulations, $\epsilon $ is an adjustable parameter for ensuring that the matrix is invertible, which should be as small as possible.
	For example, we set $ \epsilon = 0.1$ for measurement noise setting (1).}
Under a test dataset with measurement noise setting different from the training dataset, it is difficult for NN to predict a very accurate $\hat{\mathbf{e}}$, but it can predict a relatively accurate weighting matrix $\mathbf{W}$,
which makes the NN-WLS more robust than the NN-LS.
Therefore, the robustness of the proposed NN-WLS outperforms the NN-LS and the black box NN in the given simulation scenarios.

\subsubsection{Ensemble Learning-based NN-WLS Methods}

We analyze the MAE performance of the proposed ensemble learning-based NN-WLS methods by setting $\delta_d=3$ m and $\delta_a=0.0175$ rad. The radius $r_a$ for the subtractive clustering algorithm in ENN-A-WLS is set to $0.1$, $0.01$, and $0.001$ for ratios $0.1$, $0.01$, and $0.001$, respectively.
The number of emsembled NNs is $P=100$.
The size of the training dataset for NN-WLS, ENN-M-WLS, ENN-A-WLS, and ENN-B-WLS is $12,000$.
Here, ENN-M-WLS has the same structure as ENN-A-WLS but uses a simpler averaging method to replace the subtractive clustering algorithm in ENN-A-WLS.
The MAE results are presented in Table \ref{tab:ensemble_u} and Table \ref{tab:ensemble_v}.
The proposed ENN-B-WLS has the best performance in terms of location estimation.
The reason is that the combination of the predictions of multiple NNs by \eqref{bennwls} can approximate the statistical characteristic of $\mathbf{W}$ remarkably.
However, the ensemble learning-based NN-WLS method has no evident advantages over NN-WLS in terms of velocity estimation.
Not much space for improvement is needed because the values of the velocity in simulations are relatively small, indicating that the estimation error of NN-WLS is also small.

\begin{table}
	\vspace{1.5cm}
	\centering
	\begin{minipage}[t]{0.48\textwidth}
		\renewcommand{\arraystretch}{1.5}
		\centering
		\fontsize{8}{8}\selectfont
		\captionsetup{font=small}
		\begin{threeparttable}
			\caption{MAE (m) performance comparison of UE location estimation.}
			\label{tab:ensemble_u}
			\begin{tabular}{cccccccccc}
				\toprule
				\multirow{2}{*}{Method}&
				\multicolumn{3}{c}{ $\delta_d=3 $ m, $\delta_a=0.0175$ rad}\cr
				\cmidrule(lr){2-4} 
				&Ratio=0.1&Ratio=0.01&Ratio=0.001\cr
				\midrule
				WLS & 3.86 & 3.85 & 3.85 \cr			
				NN-WLS &  2.24 &  0.55 &  0.09 \cr	
				ENN-M-WLS & 2.41 & 0.53 & 0.13 \cr
				ENN-A-WLS & 1.99 & 0.53 & 0.11 \cr
				ENN-B-WLS & {\bf0.60} & {\bf0.24} & {\bf0.05} \cr					
				\bottomrule
			\end{tabular}
		\end{threeparttable}
	\end{minipage}
	\hspace{0.15cm}
	\begin{minipage}[t]{0.48\textwidth}
		\renewcommand{\arraystretch}{1.5}
		\centering
		\fontsize{8}{8}\selectfont
		\captionsetup{font=small}
		\begin{threeparttable}
			\caption{MAE (m/s) performance comparison of UE velocity estimation.}
			\label{tab:ensemble_v}
			\begin{tabular}{cccccccccc}
				\toprule
				\multirow{2}{*}{Method}&
				\multicolumn{3}{c}{ $\delta_d=3 $ m, $\delta_a=0.0175$ rad}\cr
				\cmidrule(lr){2-4} 
				&Ratio=0.1&Ratio=0.01&Ratio=0.001\cr
				\midrule
				WLS & 2.88 & 2.80 & 2.79 \cr				
				NN-WLS & {\bf 0.59} & 0.15 & {\bf 0.13} \cr		
				ENN-M-WLS & 0.59 & 0.20 & 0.17 \cr
				ENN-A-WLS & {\bf 0.59} & {\bf 0.14} & {\bf 0.13} \cr
				ENN-B-WLS & 0.71 & 0.18 & 0.15 \cr		
				\bottomrule
			\end{tabular}
		\end{threeparttable}
	\end{minipage}
	\vspace{-1cm}
\end{table}

\subsubsection{Time Resources}
We compare the time resources consumed by different localization methods. 
The model-based WLS estimator needs $0.06$ seconds when executed on a desktop computer with a 3.3 GHz Intel(R) Xeon(R) W-2155 CPU and 64 GB of RAM, using Windows 10 and MATLAB 2018b (64-bit).
The time needed by the NN-WLS consists of two parts.
The test time of the NN is $1.6 \times 10^{-7}$ seconds when tested on the 1080 Ti GPU,
and the time required to input the results of the NN into the WLS estimator and obtain the final estimation is $0.011$ seconds when using MATLAB.
Thus, the total time needed by the NN-WLS is $0.011$ seconds, which is $17\%$ of the time needed by the model-based WLS estimator.
The ensemble learning-based NN-WLS takes more time than the NN-WLS because the ensemble time is $1.6 \times 10^{-3}$ seconds when using MATLAB.
Thus, the total time needed by the ensemble learning based NN-WLS is $0.013$ seconds,
which is $22\%$ of the time needed by the model-based WLS estimator.
This result is reasonable because the WLS algorithm requires initialization and several update processes, which are time consuming, whereas the NN-WLS and the ensemble learning based NN-WLS do not need to execute such processes.

\section{Conclusion}
This study considered the joint location and velocity estimation problem in a 3-D mmWave CRAN architecture. First, we embedded the cooperative localization into communications and established the joint location and velocity estimation model with hybrid TDOA/FDOA/AOA measurements. Then, an efficient closed-form WLS solution, which was subsequently proven asymptotically unbiased under small noise levels, was deduced. Second, we built the scatterer localization model by exploiting the single-bounce NLOS paths and the estimated UE location and deduced the closed-form WLS solution.  
The simulation results indicated that the WLS-based joint estimation algorithm can achieve the CRLB and outperform the benchmarks.

Furthermore, the NN-WLS algorithm was proposed by embedding the NNs into the proposed WLS estimators to replace linear approximation. 
This study is the first to combine the WLS estimator and NN in 3-D localization methods in the existing literature. 
The combination harnesses both powerful learning ability of the NN and the robustness of the proposed geometric model. 
In addition, ensemble learning was introduced to improve performance. 
A revised ray-tracing dataset was used in the simulations to test the performance of the NN-WLS algorithm. 
Simulation results showed that NN-WLS is fast because it can eliminate iterations in the proposed WLS algorithm, 
and significantly outperforms the WLS algorithm when the measurement error vector exhibits some correlation pattern.
In addition, through a comprehensive comparison with the black box NN and the NN-LS method, the proposed NN-WLS is more excellent in terms of localization accuracy and robustness.

\begin{appendices}
\section{}\label{B}
In this section, we approximate $\mathbf{e}$ up to the linear noise term in \eqref{e}.
For the differentiable function $f(x_1,\ldots,x_n)$ on the variables $x_1,\ldots,x_n$,
there holds
\vspace{-0.25cm}
\begin{equation}\label{23}
f(x_1+\Delta x_1,\ldots,x_n+\Delta x_n)-f(x_1,\ldots,x_n) =
\frac{\partial f}{\partial x_1}\Delta x_1
+\ldots+\frac{\partial f}{\partial x_n}\Delta x_n+o(\eta) ,
\vspace{-0.25cm}
\end{equation}
where $\eta=\sqrt{(\Delta x_1)^2+\ldots+(\Delta x_n)^2}\rightarrow 0$.
According to \eqref{191} and \eqref{20}, we get
\vspace{-0.25cm}
\begin{equation}\label{e0}
\mathbf{e}=(\tilde{\mathbf{h}}-\tilde{\mathbf{G}}\mathbf{x}^{\circ})-(\mathbf{h}-\mathbf{G}\mathbf{x}^\circ).
\vspace{-0.25cm}
\end{equation}
Applying (\ref{23}) with (\ref{e0}), firstly, for $i=2,\ldots,N_{a}$, we yield
the $(2i\!-\!3)$-th entry in $\mathbf{e}$ as
{\begingroup\makeatletter\def\f@size{11}\check@mathfonts
	\def\maketag@@@#1{\hbox{\m@th\normalsize\normalfont#1}}\setlength{\arraycolsep}{0.0em}\setlength{\arraycolsep}{0.0em}
	\begin{eqnarray*}\label{24}
\begin{aligned}
\hspace{-0.14cm}
\mathbf{e}(2i\!-\!3)
&\!\!\approx\!\![2r_{i1}^{\circ}\!\!+\!\!2\mathbf{a}_1^{\circ T}\!(\mathbf{u}^{\circ}\!\!-\!\mathbf{b}_1)]\Delta r_{i1}
\!\!+\!\!\left[\!-2r_{i1}^{\circ}\frac{\partial \mathbf{a}_1^{\circ T}}{\partial{\phi_1^{\circ}}}\mathbf{b}_1
\!\!+\!\!2r_{i1}^{\circ}\frac{\partial \mathbf{a}_1^{\circ T}}{\partial{\phi_1^{\circ}}}\mathbf{u}^{\circ}\right]\!\!
\Delta\phi_1
\!\!+\!\!\left[\!-2r_{i1}^{\circ}\frac{\partial \mathbf{a}_1^{\circ T}}{\partial{\theta_1^{\circ}}}\mathbf{b}_1
+2r_{i1}^{\circ}\frac{\partial \mathbf{a}_1^{\circ T}}{\partial{\theta_1^{\circ}}}\mathbf{u}^{\circ}\right]\!\!
\Delta\theta_1,
\end{aligned}
\end{eqnarray*}\setlength{\arraycolsep}{5pt}\endgroup}where $\frac{\partial \mathbf{a}_1^{\circ T}}{\partial{\phi_1^{\circ}}}(\mathbf{u}^{\circ}-\mathbf{b}_1)= 0$,
$\frac{\partial \mathbf{a}_1^{\circ T}}{\partial{\theta_1^{\circ}}}(\mathbf{u}^{\circ}-\mathbf{b}_1)= 0$, and $\mathbf{a}_1^{\circ T}(\mathbf{u}^{\circ}-\mathbf{b}_1) = r_1^{\circ}$, hence, we have
\vspace{-0.25cm}
\begin{equation}\label{241}
\mathbf{e}(2i-3)\approx2r_i^{\circ} \Delta r_{i1}.
\vspace{-0.4cm}
\end{equation}
Similarly, we have
\vspace{-0.25cm}
\begin{equation}\label{25i}
\mathbf{e}(2i-2)\approx \dot{r}_{i}^{\circ}\Delta r_{i1}+r_{i}^{\circ}\Delta \dot{r}_{i1}+r_{1}^{\circ}r_{i1}^{\circ}\cos^{2}\theta_1^{\circ}\dot{\phi}_1^\circ\Delta\phi_1
+r_{1}^{\circ}r_{i1}^{\circ}\dot{\theta}_1^\circ\Delta\theta_1.
\vspace{-0.25cm}
\end{equation}
For $j=1,\ldots,N_{a}$, we have
\vspace{-0.25cm}
\begin{equation}\label{26}
\mathbf{e}(2N_{a}-3+2j)\approx\left(\frac{\partial{\mathbf{c}_j^{\circ T}}}{\partial{\phi_j^{\circ}}}\mathbf{b}_j
-\frac{\partial{\mathbf{c}_j^{\circ T}}}
{\partial{\phi_j^{\circ}}}\mathbf{u}^{\circ}\right)\Delta\phi_j=r_j^{\circ}\cos\theta_j^{\circ}\Delta\phi_j.
\vspace{-0.25cm}
\end{equation}
and
\vspace{-0.25cm}
\begin{equation}\label{28}
\begin{aligned}
\mathbf{e}(2N_{a}-2+2j)&\approx\frac{\partial{\mathbf{d}_j^{\circ T}}}{\partial{\phi_j^{\circ}}}
\left(\mathbf{b}_j
-\mathbf{u}^{\circ}\right)\Delta\phi_j +\frac{\partial{\mathbf{d}_j^{\circ T}}}{\partial{\theta_j^{\circ}}}
\left(\mathbf{b}_j
-\mathbf{u}^{\circ}\right)\Delta\theta_j=r_j^{\circ}\Delta\theta_j.
\end{aligned}
\vspace{-0.25cm}
\end{equation}
Finally, transforming the expressions \eqref{241}, \eqref{25i}, \eqref{26}, and \eqref{28} for $i=2,\ldots,N_{a}$ and $j=1,\ldots,N_{a}$ into matrix representation, we obtain the first-order approximation of $\mathbf{e}$ as
$\mathbf{e} \approx \mathbf{B}\Delta \mathbf{m}$ in \eqref{e}.

\section{}\label{D}
In this section, we take the state of UE $\mathbf{x}^\circ$ as an example.
We first calculate the partial derivatives required for CRLB. 	
	According to \cite{KAY}, the CRLB of $\mathbf{x}^{\circ}$ for the Gaussian noise model can be defined as
	\vspace{-0.25cm}
	\begin{equation}\label{b1}
	\mbox{\mbox{CRLB}}(\mathbf{x}^{\circ})=(\mathbf{D}^T\mathbf{Q}^{-1}\mathbf{D})^{-1},
	\vspace{-0.25cm}
	\end{equation}
	where $\mathbf{D}=\partial \mathbf{m}^\circ/ \partial {\mathbf{x}^{\circ T}}$.
	The partial derivatives are given by
	\begin{equation}\label{335}	
	\hspace{-0.25cm}
	\frac{\partial \mathbf{m}^\circ}{\partial\mathbf{x}^{\circ T}}
	\!\!=\!\!\left[\!(\frac{\partial r_{21}^{\circ }}{\partial\mathbf{x}^{\circ T}})^T,
	(\frac{\partial \dot{r}_{21}^{\circ }}{\partial\mathbf{x}^{\circ T}})^T,\ldots,
	(\frac{\partial r_{N_{a} 1}^{\circ }}{\partial\mathbf{x}^{\circ T}})^T,
	(\frac{\partial \dot{r}_{N_{a} 1}^{\circ }}{\partial\mathbf{x}^{\circ T}})^T,
	(\frac{\partial \phi_{1}^{\circ }}{\partial\mathbf{x}^{\circ T}})^T,
	(\frac{\partial \theta_{1}^{\circ }}{\partial\mathbf{x}^{\circ T}})^T,\ldots,
	(\frac{\partial \phi_{N_{a}}^{\circ }}{\partial\mathbf{x}^{\circ T}})^T,
	(\frac{\partial \theta_{N_{a}}^{\circ }}{\partial\mathbf{x}^{\circ T}})^T
	\!\right]^T\!\!\!\!,
	\vspace{-0.25cm}
	\end{equation}
	and 
	\begin{equation}\label{35}
	\hspace{-0.35cm}
	\frac{\partial r_{i1}^{\circ }}{\partial\mathbf{x}^{\circ T}}\!=\!
	\left[\frac{\partial r_{i1}^{\circ }}{\partial \mathbf{u}^{\circ T}},
	\frac{\partial r_{i1}^{\circ }}{\partial \dot{\mathbf{u}}^{\circ T}}\right],
	\frac{\partial \dot{r}_{i1}^{\circ }}{\partial\mathbf{x}^{\circ T}}\!=\!
	\left[\frac{\partial \dot{r}_{i1}^{\circ }}{\partial \mathbf{u}^{\circ T}},
	\frac{\partial \dot{r}_{i1}^{\circ }}{\partial \dot{\mathbf{u}}^{\circ T}}\right],
	\frac{\partial \phi_{j}^{\circ }}{\partial\mathbf{x}^{\circ T}}\!=\!
	\left[\frac{\partial \phi_{j}^{\circ }}{\partial \mathbf{u}^{\circ T}},
	\frac{\partial \phi_{j}^{\circ }}{\partial \dot{\mathbf{u}}^{\circ T}}\right],
	\frac{\partial \theta_{j}^{\circ }}{\partial\mathbf{x}^{\circ T}}\!=\!
	\left[\frac{\partial \theta_{j}^{\circ }}{\partial \mathbf{u}^{\circ T}},
	\frac{\partial \theta_{j}^{\circ }}{\partial \dot{\mathbf{u}}^{\circ T}}\right]\!\!,
	\end{equation}
	where $i=2,\ldots,N_{a}$ and $j=1,\ldots,N_{a}$.
Firstly, from (\ref{1}) and (\ref{2}), we obtain
\vspace{-0.25cm}
\begin{equation}\label{p1}
\dfrac{\partial r_{i1}^{\circ}}{\partial \mathbf{u}^{\circ T}}=\dfrac{(\mathbf{u}^{\circ}-\mathbf{b}_i)^T}{r_i^{\circ}}-\dfrac{(\mathbf{u}^{\circ}-\mathbf{b}_1)^T}{r_1^\circ}, \quad \dfrac{\partial r_{i1}^{\circ}}{\partial \dot{\mathbf{u}}^{\circ T}}=\mathbf{0}.
\vspace{-0.25cm}
\end{equation}
Secondly, from (\ref{3}) and (\ref{4}), we get
\vspace{-0.25cm}
\begin{equation}\label{p2}
\dfrac{\partial\dot{r}_{i1}^{\circ}}{\partial \mathbf{u}^{\circ T}}=\dfrac{\dot{r}_1^{\circ}(\mathbf{u}^{\circ}-\mathbf{b}_1)^T}{(r_1^{\circ })^2}-\dfrac{\dot{r}_i^{\circ}(\mathbf{u}^{\circ}-\mathbf{b}_i)^{T}}{(r_i^{\circ })^2}+\dfrac{\dot{\mathbf{u}}^{\circ T}}{r_i^{\circ}}-\dfrac{\dot{\mathbf{u}}^{\circ T}}{r_1^{\circ}}, \ \ \
\dfrac{\partial\dot{r}_{i1}^{\circ}}{\partial \dot{\mathbf{u}}^{\circ T}}=\dfrac{(\mathbf{u}^{\circ}-\mathbf{b}_i)^T}{r_i^{\circ}}-\dfrac{(\mathbf{u}^{\circ}-\mathbf{b}_1)^T}{r^{\circ}_1}.
\vspace{-0.25cm}
\end{equation}
Thirdly, according to (\ref{17}), we have
$
(\mathbf{b}_j-\mathbf{u}^{\circ})^T{\partial \mathbf{c}^{\circ}_j}/{\partial \mathbf{u}^{\circ T}}=\mathbf{c}^{\circ T}_j.
$
Since $\mathbf{a}_j^{\circ T}[\cos\phi^{\circ}_j,\sin\phi^{\circ}_j,0]^T=\cos\theta^{\circ}_j$, we yield
$	(\mathbf{b}_j -\mathbf{u}^{\circ})^T{\partial \mathbf{c}^{\circ}_j}/{\partial \mathbf{u}^{\circ T}} =-r^{\circ}_j\mathbf{a}_j^{\circ T}\!{\partial \mathbf{c}^{\circ}_j}/{\partial \mathbf{u}^{\circ T}}= r^{\circ}_j\!\cos\theta_j^{\circ}{\partial \phi^{\circ}_j}/{\partial \mathbf{u}^{\circ T}},
$
that is,
\vspace{-0.25cm}
\begin{equation}\label{p3}
\dfrac{\partial \phi^{\circ}_j}{\partial \mathbf{u}^{\circ T}}=\dfrac{\mathbf{c}^{\circ T}_j}{r^{\circ}_j\cos\theta_j^{\circ}},\ \
\dfrac{\partial \phi^{\circ}_j}{\partial \dot{\mathbf{u}}^{\circ T}}=\mathbf{0},
\vspace{-0.25cm}
\end{equation}
for $j=1,\ldots,N_{a}$.
Similarly, from \eqref{17}, we obtain
$
(\mathbf{u}^{\circ} \!-\!\mathbf{b}_j)^T{\partial \mathbf{d}^{\circ}_j}/{\partial \mathbf{u}^{\circ T}}+\mathbf{d}^{\circ T}_j=\mathbf{0},
$
that is,
$
(\mathbf{u}^{\circ} \!-\!\mathbf{b}_j)^T[\dfrac{\partial  \mathbf{d}^{\circ}_j}{\partial \theta^{\circ}_j}\dfrac{\partial\theta^{\circ}_j}{\partial \mathbf{u}^{\circ T}}+\dfrac{\partial \mathbf{d}^{\circ}_j}{\partial \phi^{\circ}_j}\dfrac{\partial\phi^{\circ}_j}{\partial \mathbf{u}^{\circ T}}]=-\mathbf{d}^{\circ T}_j.
$
Since $(\mathbf{u}^{\circ} \!-\!\mathbf{b}_j)^T=r^{\circ}_j\mathbf{a}^{\circ T}_j$,
$\mathbf{a}^{\circ T}_j{\partial \mathbf{d}^{\circ}_j}/{\partial \theta^{\circ}_j}=-1$ and
$ \mathbf{a}^{\circ T}_j{\partial \mathbf{d}^{\circ}_j}/{\partial \phi^{\circ}_j}=0$,
we get
\vspace{-0.25cm}
\begin{equation}\label{p5}
\dfrac{\partial \theta^{\circ}_j}{\partial \mathbf{u}^{\circ T}}=\dfrac{\mathbf{d}^{\circ T}_j}{r^{\circ}_j},\ \
\dfrac{\partial \theta^{\circ}_j}{\partial \dot{\mathbf{u}}^{\circ T}}=\mathbf{0}.
\vspace{-0.25cm}
\end{equation}

Next, we prove that $\mbox{cov}(\mathbf{x})\approx\mbox{CRLB}(\mathbf{x}^{\circ})$ under small noise levels.
The proof relies on the following two key identities, for $i=2,\ldots,N_{a}$,
\begin{eqnarray}
(a):&& r_{i}^{\circ}\left[\frac{(\mathbf{u}^{\circ}-\mathbf{b}_{i})^T}{r_{i}^{\circ}}-\frac{(\mathbf{u}^{\circ}-\mathbf{b}_{1})^{T}}{r_{1}^{\circ}}\right]=(\mathbf{b}_{1}-\mathbf{b}_{i})^{T}-r_{i1}^{\circ}\mathbf{a}_{1}^{\circ T},\label{aa}\\
(b):&& \dot{r}_{i}^{\circ}\left[\frac{(\mathbf{u}^{\circ}-\mathbf{b}_{i})^{T}}{r_{i}^{\circ}}-\frac{(\mathbf{u}^{\circ}-\mathbf{b}_{1})^{T}}{r_{1}^{\circ}}\right]+r_{i}^{\circ}\bigg[\frac{\dot{r}_{1}^{\circ}(\mathbf{u}^{\circ}-\mathbf{b}_{1})^{T}}{(r_{1}^{\circ})^{2}}\!
-\!\frac{\dot{r}_{i}^{\circ}(\mathbf{u}^{\circ}-\mathbf{b}_{i})^{T}}{(r_{i}^{\circ})^{2}}
\!+\!\frac{\dot{\mathbf{u}}^{\circ T}}{r_{i}^{\circ}}\!-\!\frac{\dot{\mathbf{u}}^{\circ T}}{r_{1}^{\circ}}\bigg]
 \nonumber \\
&&+r_{i1}^{\circ}\dot{\phi}_{1}^{\circ}\cos\theta_{1}^{\circ}\mathbf{c}_{1}^{\circ T}+r_{i1}^{\circ}\dot{\theta}_{1}^{\circ}\mathbf{d}_{1}^{\circ T}= -\dot{r}_{i1}^{\circ}\mathbf{a}_{1}^{\circ T}. \label{bb}
\vspace{-0.25cm} 
\end{eqnarray}
Since $(\mathbf{u}^{\circ} \!-\!\mathbf{b}_j)^T=r^{\circ}_j\mathbf{a}^{\circ T}_j$,
$\dot{\mathbf{u}}^\circ=\dot{r}_1^\circ \mathbf{a}_1^\circ+r_1^\circ\dot{\mathbf{a}}_1^\circ$, and $\dot{\phi}_{1}^{\circ}{\partial \mathbf{a}_{1}^{\circ T}}/{\partial\phi_{1}^{\circ}}+\dot{\theta}_{1}^{\circ}{\partial \mathbf{a}_{1}^{\circ T}}/{\partial\theta_{1}^{\circ}}=\dot{\mathbf{a}}_{1}^{\circ T}$,
by some tedious derivation, we can prove that (a) and (b) hold.

\end{appendices}

\end{document}